\documentclass[11pt,reqno,oneside]{article}
\usepackage{pstricks,pst-node,pst-tree}
\usepackage{amsmath}
\usepackage{amssymb}
\usepackage[lmargin=.55in, rmargin=.55in, tmargin=1in, bmargin=1in]{geometry}
\usepackage{xcolor}
\usepackage[dvips]{graphicx}
\usepackage{pstricks}

\newtheorem{lemm}{Lemma}
\newtheorem{prop}{Proposition}
\newtheorem{coro}{Corollary}

\definecolor{myblue}{rgb}{0,0.1,0.9}
\definecolor{myred}{rgb}{0.9,0.1,0}

\title{Asymptotic properties of the number of matching coalescent histories for caterpillar-like families of species trees}
\author{Filippo Disanto\thanks{Corresponding author. Email: fdisanto@stanford.edu.}
\\ Noah A.~Rosenberg \\
\\ {\small Department of Biology, Stanford University, Stanford, CA 94305 USA} }

\begin{document}

\maketitle

\begin{abstract}
Coalescent histories provide lists of species tree branches on which gene tree coalescences can take place, and their enumerative properties assist in understanding the computational complexity of calculations central in the study of gene trees and species trees. Here, we solve an enumerative problem left open by Rosenberg (\emph{IEEE/ACM Transactions on Computational Biology and Bioinformatics} 10: 1253-1262, 2013) concerning the number of coalescent histories for gene trees and species trees with a matching labeled topology that belongs to a generic caterpillar-like family. By bringing a generating function approach to the study of coalescent histories, we prove that for any caterpillar-like family with seed tree $t$, the sequence $(h_n)_{n\geq 0}$ describing the number of matching coalescent histories of the $n$th tree of the family grows asymptotically as a constant multiple of the Catalan numbers.  Thus, $h_n \sim \beta_t c_n$, where the asymptotic constant $\beta_t > 0$ depends on the shape of the seed tree $t$. The result extends a claim demonstrated only for seed trees with at most 8 taxa to arbitrary seed trees, expanding the set of cases for which detailed enumerative properties of coalescent histories can be determined. We introduce a procedure that computes from $t$ the constant $\beta_t$ as well as the algebraic expression for the generating function of the sequence $(h_n)_{n\geq 0}$.
\end{abstract}

%%%%%%%%%%%%%%%%%%%%%%%%%%%%%%%%%%%%%%%%%%%%%%%%%%%%%%%%%
%%%%%%%%%%%%%%%%%%%%%%%%%%%%%%%%%%%%%%%%%%%%%%%%%%%%%%%%%
%%%%%%%%%%%%%%%%%%%%%%%%%%%%%%%%%%%%%%%%%%%%%%%%%%%%%%%%%

\section{Introduction}
\label{secIntro}

Coalescent histories, mathematical structures representing combinatorially distinct ways in which a given gene tree can coalesce along the branches of a given species tree, are important in a variety of phylogenetic problems~\cite{DisantoAndRosenberg15, Rosenberg13:tcbb, RosenbergAndDegnan10}. They arise most prominently in characterizing the set of objects over which a sum is performed in a fundamental calculation for inference of species trees from information on multiple genetic loci, the evaluation of gene tree probabilities conditional on species trees~\cite{DegnanAndSalter05}.

Because of the appearance of coalescent histories in sets over which sums are computed, as well as in state spaces of certain phylogenetic Markov chains~\cite{DutheilEtAl09, HobolthEtAl07, HobolthEtAl11}, solutions to enumerative problems involving coalescent histories contribute to an understanding of the computational complexity of phylogenetic calculations. A recursion for the number of coalescent histories for a given gene tree and species tree has been established~\cite{Rosenberg07:jcb}, and several studies have reported exact numerical results and closed-form expressions for the number of coalescent histories for small trees and for specific types of trees of arbitrarily large size~\cite{Degnan05, DegnanAndSalter05, DisantoAndRosenberg15, Rosenberg07:jcb, Rosenberg13:tcbb, RosenbergAndDegnan10, ThanEtAl07}. The latter computations have proceeded both by solving or deploying the recursion in specific cases~\cite{Rosenberg07:jcb, Rosenberg13:tcbb, RosenbergAndDegnan10, ThanEtAl07}, as well as by identifying correspondences between coalescent histories and other combinatorial structures for which enumerative results have already been established~\cite{Degnan05, DegnanAndSalter05, DisantoAndRosenberg15}.

\begin{figure}
\begin{center}
\includegraphics*[scale=.86,trim=0 0 0 0]{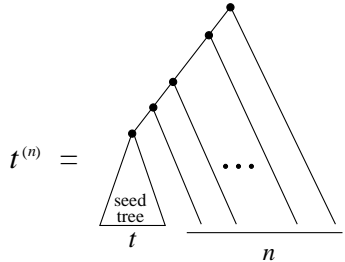}
\end{center}
\vspace{-.7cm}
\caption{{\small A caterpillar-like family of species trees $(t^{(n)})_{n\geq 0}$. For a seed tree $t$, by adding $n \geq 0$ branches each with 1 leaf, we obtain the $n$th tree of the family, $t^{(n)}$. If $t$ has 2 taxa, then $(t^{(n)})_{n\geq 0}$ is simply the caterpillar family.}}
\label{figFamilies2a1}
\end{figure}

One class of gene trees and species trees of particular interest for enumeration of coalescent histories is the \emph{caterpillar-like families}, trees that have a caterpillar shape, except that the caterpillar subtree with $r$ taxa is replaced by a subtree of size $r$ that is not necessarily a caterpillar subtree (Fig.~\ref{figFamilies2a1}). For the simplest caterpillar-like family, the set of caterpillar trees themselves, if the gene tree and species tree have the same caterpillar labeled topology with $n$ taxa, then the number of coalescent histories is a Catalan number,
\begin{equation}
\label{catalano1}
c_{n-1} = \frac{1}{n} {2n-2 \choose n-1}.
\end{equation}

For $T_r$-caterpillar-like families, in which the $r$-taxon subtree of an $n$-taxon caterpillar species tree is replaced by an $r$-taxon subtree $T_r$ (Fig.~\ref{figFamilies2a1}), by employing the recursion method, Rosenberg \cite{Rosenberg13:tcbb} obtained the exact number of coalescent histories for all $n$, for each $T_r$ with $r \leq 8$, in the case that the gene tree and species tree have the same labeled topology. Rosenberg \cite{Rosenberg13:tcbb} argued that in each of these cases, as $n \rightarrow \infty$, the number of coalescent histories is asymptotic to a constant multiple of the Catalan numbers. A proof of this result has been presented in full for each case with $r\leq 5$~\cite{Degnan05, Rosenberg07:jcb, Rosenberg13:tcbb}, and by computer algebra for cases with $r=6$, 7, and 8~\cite{Rosenberg13:tcbb}.

%%%%%%%%%%%%%%%%%%%%%%%%%%%%%%%%%%%%%%%%%%%%%%%%%%%%%%%%%
%%%%%%%%%%%%%%%%%%%%%%%%%%%%%%%%%%%%%%%%%%%%%%%%%%%%%%%%%
%%%%%%%%%%%%%%%%%%%%%%%%%%%%%%%%%%%%%%%%%%%%%%%%%%%%%%%%%
%%%%%%%%%%%%%%%%%%%%%%%%%%%%%%%%%%%%%%%%%%%%%%%%%%%%%%%%%
%%%%%%%%%%%%%%%%%%%%%%%%%%%%%%%%%%%%%%%%%%%%%%%%%%%%%%%%%
%%%%%%%%%%%%%%%%%%%%%%%%%%%%%%%%%%%%%%%%%%%%%%%%%%%%%%%%%

Here, using a substantially different approach that brings to studies of coalescent histories the methods of analytic combinatorics, we produce an enumeration result that covers caterpillar-like families in general. We show that the result of \cite{Rosenberg13:tcbb} applies to all caterpillar-like families, not only those for which $T_r$ has $r \leq 8$. That is, we demonstrate that for any $T_r$, as $n \rightarrow \infty$, the number of coalescent histories in the $T_r$-caterpillar-like family is asymptotic to a constant multiple of the Catalan numbers. We describe a method for computing the constant and provide a symbolic tool for performing the computation. Finally, we discuss the results in terms of their impact in mathematical phylogenetics.

% \fil{Next, we exhibit a family of species trees -- that we term \emph{lodgepole family} --  whose number of coalescent histories grows faster than
% exponential in the number of taxa $n$. Previous results considered only families of species trees with an exponential growth.} We then address a problem left % open by
% \cite{noah1} on the variability at a given $n$ of the number of coalescent histories for cases with matching gene trees and species trees. Rosenberg
% \cite{noah1} obtained a lower bound on the ratio of the largest number of coalescent histories to the smallest number of coalescent histories, showing that
% this ratio was greater than a constant multiple of $n \sqrt{n}$. Here
% , by consideration of a family of species trees that we term \emph{lodgepole families}, \fil{whose number of coalescent histories grows faster than
% exponential in $n$,}
% we improve substantially upon this lower bound, demonstrating that it exceeds the much larger  $\left( \frac{\sqrt{n-1}}{4\sqrt{e}}  \right)^{n}.$

%%%%%%%%%%%%%%%%%%%%%%%%%%%%%%%%%%%%%%%%%%%%%%%%%%%%%%%%%
%%%%%%%%%%%%%%%%%%%%%%%%%%%%%%%%%%%%%%%%%%%%%%%%%%%%%%%%%
%%%%%%%%%%%%%%%%%%%%%%%%%%%%%%%%%%%%%%%%%%%%%%%%%%%%%%%%%

\section{Preliminaries}
\label{secPreliminaries}

%%%%%%%%%%%%%%%%%%%%%%%%%%%%%%%%%%%%%%%%%%%%%%%%%%%%%%%%%
%%%%%%%%%%%%%%%%%%%%%%%%%%%%%%%%%%%%%%%%%%%%%%%%%%%%%%%%%
%%%%%%%%%%%%%%%%%%%%%%%%%%%%%%%%%%%%%%%%%%%%%%%%%%%%%%%%%

\subsection{Species trees and coalescent histories}
\label{secSpeciesTrees}

We consider binary rooted leaf-labeled species trees, taking a single arbitrary labeling (without loss of generality) to represent a given unlabeled species tree topology. We consider an arbitrarily labeled species tree and its unlabeled tree interchangeably, treating the labeling as implicit.

\begin{figure}
\begin{center}
\includegraphics*[scale=.76,trim=0 0 0 0]{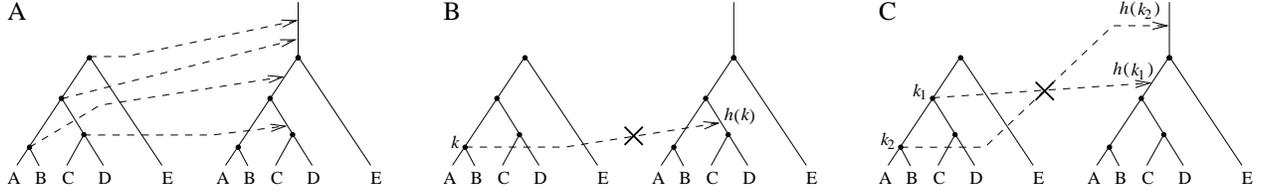}
\end{center}
\vspace{-.7cm}
\caption{{\small Matching coalescent histories. (A) A matching coalescent history. (B) A mapping from the internal nodes of a tree to its branches that does not satisfy condition (a). Leaf B is descended from node $k$ but does not descend from branch $h(k)$. (C) A mapping from the internal nodes of a tree to its internal branches that does not satisfy condition (b). Node $k_2$ is descended from node $k_1$, but branch $h(k_2)$ is strictly ancestral to branch $h(k_1)$.}}
\label{figPappone2}
\end{figure}

We examine coalescent histories for the case in which gene trees and species trees have the same labeled topology $t$, terming a coalescent history in this case a \emph{matching coalescent history}. To be a matching coalescent history, a mapping $h$ from the internal nodes of $t$ (viewed as the gene tree) to the branches of $t$ (viewed as the species tree) must satisfy two conditions: (a) for each leaf $x$ in $t$, if $x$ descends from node $k$ in $t$, then $x$ descends from branch $h(k)$ in $t$; (b) for each pair of internal nodes $k_1$ and $k_2$ in $t$, if $k_2$ descends from $k_1$ in $t$, then branch $h(k_2)$ descends from or coincides with branch $h(k_1)$ in $t$. The definition of matching coalescent histories is illustrated in Figure \ref{figPappone2}. We henceforth consider only matching coalescent histories, treating ``matching'' as implicit in references to coalescent histories; we also refer simply to \emph{histories} for short.

%%%%%%%%%%%%%%%%%%%%%%%%%%%%%%%%%%%%%%%%%%%%%%%%%%%%%%%%%
%%%%%%%%%%%%%%%%%%%%%%%%%%%%%%%%%%%%%%%%%%%%%%%%%%%%%%%%%
%%%%%%%%%%%%%%%%%%%%%%%%%%%%%%%%%%%%%%%%%%%%%%%%%%%%%%%%%
%%%%%%%%%%%%%%%%%%%%%%%%%%%%%%%%%%%%%%%%%%%%%%%%%%%%%%%%%
%%%%%%%%%%%%%%%%%%%%%%%%%%%%%%%%%%%%%%%%%%%%%%%%%%%%%%%%%
%%%%%%%%%%%%%%%%%%%%%%%%%%%%%%%%%%%%%%%%%%%%%%%%%%%%%%%%%

%%%%%%%%%%%%%%%%%%%%%%%%%%%%%%%%%%%%%%%%%%%%%%%%%%%%%%%%%
%%%%%%%%%%%%%%%%%%%%%%%%%%%%%%%%%%%%%%%%%%%%%%%%%%%%%%%%%
%%%%%%%%%%%%%%%%%%%%%%%%%%%%%%%%%%%%%%%%%%%%%%%%%%%%%%%%%

\subsection{Caterpillar-like families of species trees}
\label{secCaterpillar}

For a binary species tree $t$ with at least 2 taxa, we denote by $(t^{(n)})_{n\geq 0}$ the caterpillar-like family generated by the seed tree $t$. This family is recursively defined by taking $t^{(0)} = t$ and letting $t^{(n+1)}$ be the tree obtained by appending $t^{(n)}$ and a single leaf to a same root (Fig.~\ref{figFamilies2a1}).

Our interest is in the number of matching coalescent histories of $t^{(n)}$ for $n \geq 0$, a quantity we denote by $h_n(t)$ or simply $h_n$. We note that whereas \cite{Rosenberg13:tcbb} indexed trees by their numbers of taxa, here $n$ represents the number of taxa appended above the root of the seed tree, so that if seed tree $t$ has $|t|$ taxa, then $|t|+n$ gives the number of taxa in $t^{(n)}$.

% By using generating functions and analytic methods, we extend previous results \cite{noah2} showing that regardless of the seed tree, up to a constant
% factor, each caterpillar-like family generates asymptotically the same number of histories (\ref{catalano1}).
% This number given by the well-known \cite{stan2} \emph{Catalan} sequence.
% This work solves a problem left open in \cite{noah2} in which the asymptotic equivalence was proved by applying different techniques, but only for cases in
% which the seed tree of the caterpillar family had fewer than $9$ leaves.

% In Section~\ref{pseudo}, we study \emph{lodgepole} families $t^{[n]}$, see Fig.~\ref{families}B. We show that the number of matching histories in the family % generated by the seed tree with $1$ leaf grows as double factorial numbers. This proves the existence of certain tree-families for which the number of
% histories has grwoth rate greater than exponential with respect to the number of taxa.
% As it is shown at the end of Section~\ref{pseudo} (Eq. (\ref{ginetta})), in the lodgepole case the asymptotic equivalence holding for caterpillars is no
% longer satisfied. Indeed, the ratio between the number of histories of two different lodgepole families can diverge.

%%%%%%%%%%%%%%%%%%%%%%%%%%%%%%%%%%%%%%%%%%%%%%%%%%%%%%%%%
%%%%%%%%%%%%%%%%%%%%%%%%%%%%%%%%%%%%%%%%%%%%%%%%%%%%%%%%%
%%%%%%%%%%%%%%%%%%%%%%%%%%%%%%%%%%%%%%%%%%%%%%%%%%%%%%%%%

\subsection{Principles of analytic combinatorics}
\label{secMethods}

We rely on techniques of analytic combinatorics~\cite{FlajoletAndSedgewick09} to obtain our enumerative results, and recall several key points. In general, an integer sequence $(a_n)_{n\geq 0}$ can be associated with a formal power series $A(z) = \sum_{n=0}^\infty a_n z^n$, also termed the \emph{generating function} of the integers $a_n$. Considering $z$ as a complex variable, typically in a neighborhood of $0$, features of the function $A(z)$ are related to the growth of the coefficients $a_n$.
%In this case, the integers $a_n$ coincide with the coefficients of the Taylor expansion of $A(z)$ at $z=0$.

%, the generating function $A(z) = \sum_{n=0}^\infty a_n z^n$. Given the function $A(z)$, the coefficients $a_n$ can be extracted by considering the Taylor expansion of $A(z)$ for $z$ in a neighborhood of 0.
More precisely, generating functions, considered as complex functions, enable analyses of the asymptotic growth of the associated integer sequences through the analysis of their singularities in the complex plane. In particular, under suitable conditions, there exists a general correspondence between the singular expansion of a generating function $A(z)$ near its dominant singularities---those nearest the origin---and the asymptotic behavior of the associated coefficients $a_n$ (Chapter VI of \cite{FlajoletAndSedgewick09}). We make use of generating functions that near their unique dominant singularity can be described by means of the square root function, and for which theorems on singularity analysis of generating functions \cite{FlajoletAndSedgewick09} consequently apply.

% As a consequence for instance In particular, the exponential growth rate of the coefficients $a_n$ is dictated by the location of the dominant singularities % of $A(z)$ \cite{ancomb}. The subexponential factors of coefficients $a_n$ relate instead to the nature of the singularities (e.g. poles, branch cuts). Here
% we deal with square root singularities (see Chapter~VI of \cite{ancomb}) as the dominant singularity $z=1/4$ for the generating function $C(z)$ of Catalan
% numbers has this form (\ref{funcat}).

%%%%%%%%%%%%%%%%%%%%%%%%%%%%%%%%%%%%%%%%%%%%%%%%%%%%%%%%%
%%%%%%%%%%%%%%%%%%%%%%%%%%%%%%%%%%%%%%%%%%%%%%%%%%%%%%%%%
%%%%%%%%%%%%%%%%%%%%%%%%%%%%%%%%%%%%%%%%%%%%%%%%%%%%%%%%%

\subsection{Catalan numbers}
\label{secCatalan}

The Catalan sequence appears often in combinatorics \cite{FlajoletAndSedgewick09, GrahamEtAl94, Stanley99} and features prominently in our analysis. Rewriting eq.~(\ref{catalano1}) with index $n$ rather than $n-1$,
\begin{equation}
\label{eugenio}
c_n = \frac{1}{n+1} {{2n}\choose{n}}.
\end{equation}
The associated generating function is
\begin{equation}
\label{funcat}
C(z) = \sum_{n=0}^\infty c_n z^n = \frac{1-\sqrt{1-4z}}{2z}.
\end{equation}
By definition, if $[z^n]f(z)$ denotes the $n$th term in the power series expansion of $f(z)$ at $z=0$, we have
\begin{equation}
\label{lucia1}
c_n = [z^n]C(z) = \frac{1}{2} [z^{n+1}] (1 - \sqrt{1-4z} ) = \frac{1}{2}[z^{n+1}] (- \sqrt{1-4z} ).
\end{equation}
Asymptotically, applying Stirling's approximation $n! \sim \sqrt{2 \pi n} (n/e)^n$ to eq.~(\ref{eugenio}),
% applying Theorem~VI.1 of~\cite{FlajoletAndSedgewick09},
% \begin{equation*}
% %\label{lucia2}
% \frac{1}{2}[z^{n+1}] (-\sqrt{1-4z} ) \sim \frac{1}{2} \left[- \frac{4^{n+1} (n+1)^{-3/2}}{\Gamma(-1/2)}\right] \sim \frac{4^n}{n^{3/2}\sqrt{\pi}}.
% \end{equation*}
the Catalan sequence satisfies
\begin{equation}
\label{lucia3}
c_n \sim \frac{4^n}{n^{3/2} \sqrt{\pi}}.
\end{equation}

%%%%%%%%%%%%%%%%%%%%%%%%%%%%%%%%%%%%%%%%%%%%%%%%%%%%%%%%%
%%%%%%%%%%%%%%%%%%%%%%%%%%%%%%%%%%%%%%%%%%%%%%%%%%%%%%%%%
%%%%%%%%%%%%%%%%%%%%%%%%%%%%%%%%%%%%%%%%%%%%%%%%%%%%%%%%%

\section{The number of matching coalescent histories for caterpillar-like families}
\label{secMatching}

%%%%%%%%%%%%%%%%%%%%%%%%%%%%%%%%%%%%%%%%%%%%%%%%%%%%%%%%%
%%%%%%%%%%%%%%%%%%%%%%%%%%%%%%%%%%%%%%%%%%%%%%%%%%%%%%%%%
%%%%%%%%%%%%%%%%%%%%%%%%%%%%%%%%%%%%%%%%%%%%%%%%%%%%%%%%%

Our goal is to produce a procedure that evaluates the number of coalescent histories $h_n(t)$ for matching gene trees and species trees in the caterpillar-like family that begins with seed tree $t$, and moreover, to show that
\begin{equation}\label{equi}
h_n(t) \sim \beta_t c_n,
\end{equation}
where the multiplier $\beta_t > 0$ for the Catalan sequence is a constant depending on $t$. In other words, we wish to demonstrate that as $n \rightarrow \infty$, the ratio $h_n/c_n$ converges to a constant $\beta_t > 0$ that depends on the seed tree $t$.

First, in Section~\ref{secLowerBound}, we determine a lower bound for the number of matching coalescent histories of the $n$th tree $t^{(n)}$ of the caterpillar-like family with seed tree $t$. Next, in Section~\ref{secIterativeGeneration}, we introduce a concept of \emph{$m$-rooted histories} of a species tree $t^{(n)}$. The section provides an iterative construction of the rooted histories of $t^{(n+1)}$ from those of $t^{(n)}$, describing the construction by means of a convenient labeling scheme. We follow a commonly used combinatorial enumeration strategy \cite{BanderierEtAl02, BarcucciEtAl99} that determines a recursive succession rule for successive collections of objects in a sequence and then uses this rule to compute a generating function. In Section~\ref{secGenfun}, we use the iterative construction to produce a bivariate generating function whose coefficients $h_{n,m}$ are the numbers of $m$-rooted histories for trees $t^{(n)}$. We next obtain the generating function for the integer sequence $(h_n)_{n\geq 0}$ describing the number of matching coalescent histories for the $t^{(n)}$. Finally, using the lower bound from Section \ref{secLowerBound}, in Section \ref{secSomaro}, we apply methods of analytic combinatorics to study the asymptotic behavior of $h_n$.

\subsection{Lower bound for $h_n$}
\label{secLowerBound}

To produce a lower bound for $h_n$, we first define $\mathrm{V}$ as the tree with 2 taxa. Recalling that we index trees so that the number of taxa in a tree is $n$ more than the number of taxa in the seed tree, we have \cite{Degnan05, Rosenberg07:jcb, Rosenberg13:tcbb}
$$h_n(\mathrm{V}) = c_{n+1}.$$

\begin{figure}
\begin{center}
\includegraphics*[scale=.78,trim=0 0 0 0]{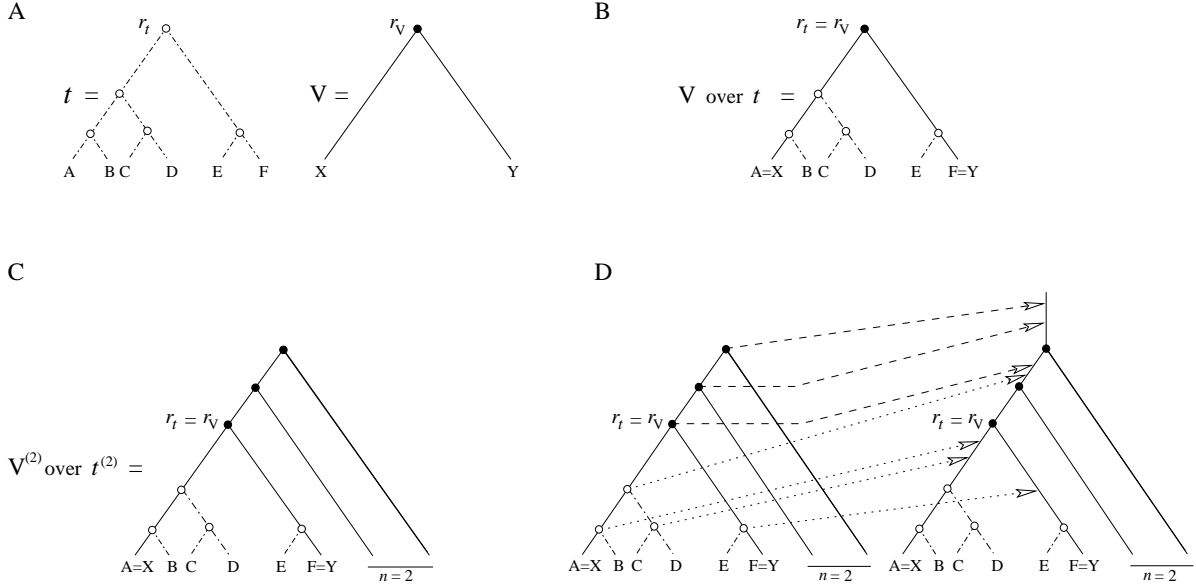}
\end{center}
\vspace{-.7cm}
\caption{{\small Superposition of the caterpillar tree family on a caterpillar-like tree family with arbitrary seed tree of size $|t| \geq 2$. (A) A seed tree $t$ and the seed tree $\mathrm{V}$ for the caterpillar family. (B) Superposition of $\mathrm{V}$ on $t$, so that the roots $r_{\mathrm{V}}$ and $r_t$ overlap. (C) Superposition of $\mathrm{V}^{(2)}$ (shaded internal nodes) on $t^{(2)}$ (shaded and unshaded nodes). The $n=2$ caterpillar branches in $\mathrm{V}^{(2)}$ and $t^{(2)}$ overlap, and $r_{\mathrm{V}}$ still matches $r_t$. (D) A matching coalescent history of $t^{(2)}$ (dashed and dotted arrows) determines a matching coalescent history of $\mathrm{V}^{(2)}$ (dashed arrows) by ignoring arrows from the unshaded nodes.
}} \label{figPalla3}
\end{figure}

A constructive procedure, illustrated in Figure \ref{figPalla3}, shows that for any seed tree $t$ with $|t| \geq 2$,
\begin{equation}\label{lov}
h_n(t) \geq h_n(\mathrm{V}) = c_{n+1}.
\end{equation}
For a seed tree $t$, we can superimpose $\mathrm{V}$ on $t$ so that the root $r_{\mathrm{V}}$ of $\mathrm{V}$ matches the root $r_t$ of $t$ (Fig.~\ref{figPalla3}B). The two leaves of $\mathrm{V}$ are identified with two of the leaves of $t$, one on each side of the root of $t$. Generating caterpillar-like families by adding $n$ single branches separately to $\mathrm{V}$ and to $t$, the superposition of $\mathrm{V}$ on $t$ extends, so that $\mathrm{V}^{(n)}$ is superimposed on $t^{(n)}$ (Fig.~\ref{figPalla3}C). The $n$ caterpillar branches of $t^{(n)}$ and $\mathrm{V}^{(n)}$ then correspond.

Each matching coalescent history $h$ of $t^{(n)}$ determines a corresponding matching coalescent history $h^\prime$ of $\mathrm{V}^{(n)}$ by considering the restriction of the history $h$ to the set of internal nodes of $t^{(n)}$ that correspond to internal nodes of $\mathrm{V}^{(n)}$ (Fig.~\ref{figPalla3}D). Thus, for any given seed tree $t$, the number of matching coalescent histories of $t^{(n)}$ is greater than or equal to the number of matching coalescent histories of $\mathrm{V}^{(n)}$. In symbols, we have eq.~(\ref{lov}).

%%%%%%%%%%%%%%%%%%%%%%%%%%%%%%%%%%%%%%%%%%%%%%%%%%%%%%%%%
%%%%%%%%%%%%%%%%%%%%%%%%%%%%%%%%%%%%%%%%%%%%%%%%%%%%%%%%%
%%%%%%%%%%%%%%%%%%%%%%%%%%%%%%%%%%%%%%%%%%%%%%%%%%%%%%%%%
%%%%%%%%%%%%%%%%%%%%%%%%%%%%%%%%%%%%%%%%%%%%%%%%%%%%%%%%%
%%%%%%%%%%%%%%%%%%%%%%%%%%%%%%%%%%%%%%%%%%%%%%%%%%%%%%%%%
%%%%%%%%%%%%%%%%%%%%%%%%%%%%%%%%%%%%%%%%%%%%%%%%%%%%%%%%%

%%%%%%%%%%%%%%%%%%%%%%%%%%%%%%%%%%%%%%%%%%%%%%%%%%%%%%%%%
%%%%%%%%%%%%%%%%%%%%%%%%%%%%%%%%%%%%%%%%%%%%%%%%%%%%%%%%%
%%%%%%%%%%%%%%%%%%%%%%%%%%%%%%%%%%%%%%%%%%%%%%%%%%%%%%%%%

\subsection{Iterative generation of rooted histories}
\label{secIterativeGeneration}

This section describes the iterative procedure that for a seed tree $t$ eventually enables us to determine a formula for $h_n$. First, in Section \ref{secRooted}, we discuss $m$-rooted histories, which extend the concept of matching coalescent histories, introducing an additional parameter $m$. Next, in Section \ref{secRelations}, we examine the relationship between rooted histories and the \emph{extended coalescent histories} of \cite{Rosenberg07:jcb}, importing results on extended coalescent histories into the more convenient framework of rooted histories. We expand our goal of enumerating matching coalescent histories for $t^{(n)}$, considering a more general problem of enumerating for $m\geq 1$ the $m$-rooted histories of $t^{(n)}$.

In Section \ref{secFoglio}, we define an operator $\Omega$ for constructing the rooted histories of $t^{(n+1)}$ from the rooted histories of $t^{(n)}$. Next, in Section \ref{secFoglietta}, we introduce a labeling scheme that in Section \ref{secAsinello} enables us to switch from counting rooted histories to counting multisets of labels. At the end of Section \ref{secIterativeGeneration}, we will have converted our enumeration problem into an enumeration that is more convenient for constructing a generating function.

\subsubsection{$m$-rooted histories}
\label{secRooted}

\begin{figure}
\begin{center}
\includegraphics*[scale=.84,trim=0 0 0 0]{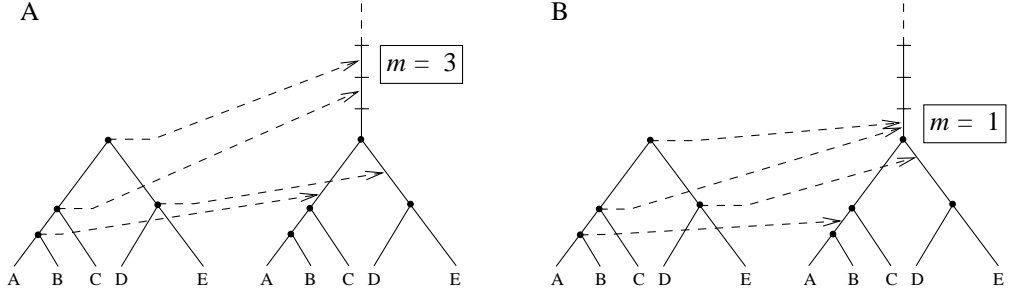}
\end{center}
\vspace{-.7cm}
\caption{{\small Rooted histories of a tree. (A) A 3-rooted history. The root-branch is divided into infinitely many components. The third component receives the image of the root. (B) A 1-rooted history. The number of 1-rooted histories corresponds to the number of matching coalescent histories of the tree. }} \label{figRooted4}
\end{figure}

Consider a tree $t$ with $|t| \geq 2$, and suppose that the branch above the root of $t$ (the \emph{root-branch}) is divided into infinitely many components. A matching coalescent history mapping the internal nodes of $t$ onto the branches of $t$ is said to be $m$-\emph{rooted} for $m \geq 1$ if the root of $t$ is mapped \emph{exactly} onto the $m$th component of the root (Fig.~\ref{figRooted4}). It is said to be \emph{rooted} if it is $m$-rooted for some $m$. Branches are numbered so that branch $m=1$ is immediately above the root node, and $m$ is greater for components that are farther from the root.

For a rooted history $h$ of a tree $t$, $m=m(h)$ denotes the component of the root-branch of $t$ that receives the image of the root of $t$. $H_{n,m}(t)$ denotes the set of $m$-rooted histories of $t^{(n)}$, and $H_n(t) = \bigcup_{m=1}^{\infty} H_{n,m}(t)$ denotes the set of its rooted histories. The number of $m$-rooted histories of $t^{(n)}$ is $h_{n,m} = |H_{n,m}|$, and the number of 1-rooted histories $h_n = h_{n,1}$ is also the number of matching coalescent histories. Enumeration of the matching coalescent histories of $t^{(n)}$ is equivalent to enumeration of the 1-rooted histories of $t^{(n)}$.

%%%%%%%%%%%%%%%%%%%%%%%%%%%%%%%%%%%%%%%%%%%%%%%%%%%%%%%%%
%%%%%%%%%%%%%%%%%%%%%%%%%%%%%%%%%%%%%%%%%%%%%%%%%%%%%%%%%
%%%%%%%%%%%%%%%%%%%%%%%%%%%%%%%%%%%%%%%%%%%%%%%%%%%%%%%%%
%%%%%%%%%%%%%%%%%%%%%%%%%%%%%%%%%%%%%%%%%%%%%%%%%%%%%%%%%
%%%%%%%%%%%%%%%%%%%%%%%%%%%%%%%%%%%%%%%%%%%%%%%%%%%%%%%%%
%%%%%%%%%%%%%%%%%%%%%%%%%%%%%%%%%%%%%%%%%%%%%%%%%%%%%%%%%

%%%%%%%%%%%%%%%%%%%%%%%%%%%%%%%%%%%%%%%%%%%%%%%%%%%%%%%%%
%%%%%%%%%%%%%%%%%%%%%%%%%%%%%%%%%%%%%%%%%%%%%%%%%%%%%%%%%
%%%%%%%%%%%%%%%%%%%%%%%%%%%%%%%%%%%%%%%%%%%%%%%%%%%%%%%%%

\subsubsection{Rooted histories and extended histories}
\label{secRelations}

% We import some results shown in \cite{Rosenberg07:jcb} for extended histories in the framework of rooted histories. The main result of the section is
% described in Proposition~\ref{peperoncino}. This states that for any tree $t$, the function $m \rightarrow h_{0,m}$, where the variable $m$ ranges in
% $[1,+\infty)$, is a polynomial that can be computed as the difference $h_{0,m} =  e_{t,m} - e_{t,m-1}$, by considering the number $e_{t,k}$ of $k$-extended
% histories of $t$ taken with $k \geq  0$.

Rooted histories are closely related to \emph{extended} coalescent histories, as defined by \cite{Rosenberg07:jcb}. We use this relationship to study properties of rooted histories. Rosenberg \cite{Rosenberg07:jcb} defined the set of $k$-\emph{extended} coalescent histories of a tree $t$ with $|t| \geq 1$ for integers $k\geq 1$; we also consider $k=0$ by setting the number of 0-extended histories to 0.

A $k$-extended history is defined as a coalescent history for a species tree whose root-branch is divided into exactly $k \geq 0$ parts. In other words, the root-branch has exactly $k\geq 0$ possible components onto which a $k$-extended history can map the gene tree root. Here we consider matching $k$-extended histories, so that the internal nodes of a tree $t$ are mapped to the branches of $t$ and its $k$ components above the root. For convenience, we refer to extended histories by the index $k$, reserving the index $m$ for rooted histories.

By the definitions of $k$-extended and $m$-rooted histories, for each $k \geq 0$, the set of $k$-extended histories of a tree is exactly the set of all $m$-rooted histories with $1 \leq m \leq k$. Therefore, for a tree $t$ with at least 2 leaves, if we label by $e_{t,k}$ its number of $k$-extended histories, then for each $m\geq 1$ the number of $m$-rooted histories of $t$ is
\begin{equation}\label{clinclin}
h_{0,m}=e_{t,m}-e_{t,m-1}.
\end{equation}
Note that for $m=1$, we explicitly use in eq.~(\ref{clinclin}) the fact that $e_{t,0}$ is defined and equal to 0. In addition to setting $e_{t,0} = 0$ for any tree $t$, as in \cite{Rosenberg07:jcb} we also set $e_{t,k} = 1$ for all $k \geq 1$ in the case that $t$ has exactly 1 leaf.

Suppose $|t| \geq 1$ and $k \geq 0$. Denote by $t_L$ and $t_R$ the left and right subtrees of the root of $t$. We can compute $e_{t,k}$ recursively as in Theorem 3.1 of \cite{Rosenberg07:jcb}:
\begin{equation}\label{peze}
e_{t,k} = \left\{
  \begin{array}{l l}
    0 & \text{if } |t| \geq 1 \text{ and } k=0  \\
    1 & \text{if } |t|=1 \text{ and } k \geq 1  \\
    \sum_{i=1}^{k} e_{t_L,i+1} e_{t_R,i+1} & \text{if } |t| \geq 2 \text{ and } k \geq 1.
  \end{array} \right.
\end{equation}

As was already observed in the remarks following Corollary~3.2 of \cite{Rosenberg07:jcb}, by eq.~(\ref{peze}), for any tree $t$ with $|t| \geq 1$, for positive integers $k\geq 1$, the function $f(k)=e_{t,k}$ is a polynomial in $k$. With our extension to permit $k=0$, we can extend this fact to $k\geq 0$ for $|t|\geq 2$: for any tree $t$ with $|t| \geq 2$, and for $k \geq 0$, we claim that the function $f(k)=e_{t,k}$ is a polynomial in $k$. Note that in allowing $k=0$, we claim $e_{t,k}$ is a polynomial in $k$ only for $|t| \geq 2$; for $|t|=1$, $e_{t,k}$ is not a polynomial in $k$ because $e_{t,0}=0$ and $e_{t,k}=1$ for $k\geq 1$.

% The tree with $1$ leaf is not considered in $(ii)$, because, in that case, we have $e_{t,k} = 0$ if $k=0$ and $e_{t,k}= 1$ if $k\geq 1$. Therefore, when $t$ % has only $1$ leaf, no polynomial in $k$ could compute $e_{t,k}$ for all $k \geq 0$ at the same time. On the other hand,
% In other words, what we want to show is that when $t$ has at least two leaves, we can always factor a term $k$ in the polynomial function $k \rightarrow
% e_{t,k}$ computed by the procedure (\ref{recnoah}). In this way

To prove the claim, fix $t$ with $|t|\geq 2$ and consider the variable $k$ over domain $[1,\infty)$. We demonstrate that $f(k)$ is a polynomial in $k$ for domain $[0,\infty)$ by showing that the closed-form polynomial for $f(k)$ has a factor of $k$, so that our choice $e_{t,0}=0$ in eq.~(\ref{peze}) is compatible with the polynomial expression valid for $k \geq 1$.

Observe that for $i \geq 1$, $e_{t_L,i}$ and $e_{t_R,i}$ are polynomials in $i$, say $P_{t_L}(i)$ and $P_{t_R}(i)$. Replacing the terms $e_{t_L,i+1}$ and $e_{t_R,i+1}$ that appear in the recursion in eq.~(\ref{peze}) by the two polynomials $P_{t_L}(i+1)$ and $P_{t_R}(i+1)$, we obtain
\begin{equation}\label{macis}
\sum_{i=1}^{k} e_{t_L,i+1} e_{t_R,i+1} = \sum_{i=1}^k P_{t_L}(i+1) \, P_{t_R}(i+1) =  \sum_{i=1}^k P^\prime(i),
\end{equation}
where $P^\prime(i)$ denotes a polynomial in $i$ that results from the product of $P_{t_L}(i+1)$ and $P_{t_R}(i+1)$. By Faulhaber's formula for sums of powers of integers, symbolic sums of the form $\sum_{i=1}^k i^p$ for a fixed integer $p \geq 0$ are polynomials containing a factor of $k$ in their closed forms (Section 6.5 of \cite{GrahamEtAl94})---for example, $\sum_{i=1}^k i^3 = k^2(k+1)^2/4$. Thus, because the polynomial $P^\prime(i)$ is a linear combination of terms of the form $i^p$, the closed-form expression for the sum $\sum_{i=1}^k P^\prime(i)$ appearing in eq.~(\ref{macis}) also has a factor of $k$. It therefore has a value of 0 at $k=0$.

Functions $e_{t,k}$ for trees $t$ with $1 \leq |t| \leq 9$ and $k \geq 1$
%(for $|t|=1$ in the tables we find only $e_{t,k}=1$ and not $e_{t,k}=0$ as we set in (\ref{peze}) for $k=0$)]]}
appear in Tables 1-4 of \cite{Rosenberg07:jcb}. For $|t| \geq 2$, as we have shown, these example polynomials are divisible by the variable representing the number of components of the root-branch. By eq.~(\ref{clinclin}), we immediately obtain the following result.

% In other words, we have shown that for trees $t$ with $|t| \geq 2$, by applying the recursive sum described in (\ref{peze}), we obtain a polynomial algebraic % expression in the variable $k$ that works not only for $k \geq 1$ (as considered in \cite{noah1}) but also cancels for $k=0$ yielding the correct number
% (zero) of $0$-extended histories.
% This can be shown by observing that, when $i$ ranges from $1$ to $k\geq 1$ as in the sum (\ref{peze}), by property $(i)$  the quantity $e_{t_L,i}$ (resp.
% $e_{t_R,i}$) is given by a polynomial in $i$ denoted here by $\text{poly}_{t_L}(i)$ (resp. $\text{poly}_{t_R}(i)$).
% Thus, we can replace the terms $e_{t_L,i+1}$ and $e_{t_R,i+1}$ that appear in (\ref{peze}) by $\text{poly}_{t_L}(i+1)$ and $\text{poly}_{t_R}(i+1)$. In this % way, the sum computed in (\ref{peze}) has the form $e_{t,k} = \sum_{i=1}^k \text{poly}_{t_L}(i+1) \cdot \text{poly}_{t_R}(i+1) = \sum_{i=1}^k
% \text{poly$(i)$}$, where $\text{poly}(i)$ denotes a generic polynomial in $i$. By Faulhaber's  formula \cite{conw}, we know that sums of the form
% $\sum_{i=1}^k i^p$ ($p \geq 0$ integer) always contain a factor $k$ in their closed forms and the same is thus true for the sum $\sum_{i=1}^k \text{poly}(i)$ % giving $e_{t,k}$. Due to the presence of a factor $k$ in the expression for the function $k \rightarrow e_{t,k}$ (with $k \geq 1$), we see that

\begin{prop}\label{peperoncino}
For any tree $t$ with $|t| \geq 2$ and for $m \geq 1$, the number $h_{0,m}$ of $m$-rooted histories of $t$ is a polynomial in $m$ that can be computed by the difference in eq.~(\ref{clinclin}) using $e_{t,k}$ as in eq.~(\ref{peze}).
\end{prop}

As an example of Proposition \ref{peperoncino}, consider the tree $t=((A,B),(C,D))$, identifying this arbitrary labeling with the unlabeled tree $(()())$.
By applying the recursive procedure in eq.~(\ref{peze}), we find that for $k \geq 0$, the number of $k$-extended coalescent histories for $t$ is $e_{t,k}=\frac{1}{6}k(2k^2 + 9k + 13)$ \cite{Rosenberg07:jcb}. The difference eq.~(\ref{clinclin}) yields that for $m \geq 1$ the number of $m$-rooted histories of $t$ is $h_{0,m} = e_{t,m} - e_{t,m-1} = m^2 + 2m + 1$.

%%%%%%%%%%%%%%%%%%%%%%%%%%%%%%%%%%%%%%%%%%%%%%%%%%%%%%%%%
%%%%%%%%%%%%%%%%%%%%%%%%%%%%%%%%%%%%%%%%%%%%%%%%%%%%%%%%%
%%%%%%%%%%%%%%%%%%%%%%%%%%%%%%%%%%%%%%%%%%%%%%%%%%%%%%%%%

\subsubsection{Generating rooted histories of $t^{(n+1)}$ from rooted histories of $t^{(n)}$}
\label{secFoglio}

% Our strategy to enumerate the $1$-rooted histories of $t^{(n)}$ is to examine the more general problem of counting the $m$-rooted histories of $t^{(n)}$ for % $m \geq 1$.

This section introduces an operator $\Omega$ that generates the rooted histories of $t^{(n+1)}$ from those of $t^{(n)}$. For each rooted history $h^\prime$ of $t^{(n+1)}$, there exists exactly one rooted history $h$ of $t^{(n)}$ with $h^\prime \in \Omega(h)$.
% The set $\Omega(h)$ is defined as the set of rooted histories $h'$ of $t^{(n+1)}$ such that, after removing the most external caterpillar branch in
% $t^{(n+1)}$, the consequent restriction of $h'$ coincides with the rooted history $h$ of $t^{(n)}$ (Fig.~\ref{indietro2}).
% Since we want to avoid double or missed counting of rooted histories, it is important to note that, for each rooted history $h'$ of $t^{(n+1)}$, there exists % exactly one history $h$ of $t^{(n)}$ with $h' \in \Omega(h)$.
% As a consequence, if $\Omega$ is iterated $n$ times as $\Omega[\ldots \Omega[\Omega(H_0)] \ldots]$ starting from the set $H_0$ of rooted histories of
% $t^{(0)}$, then all the rooted histories of $t^{(n)}$ are listed exactly once.
Recalling the definitions of the sets $H_{n,m}(t)$ and $H_n(t)$ of $m$-rooted and rooted histories of $t^{(n)}$, we define $\Omega$ as follows.
% This construction can be more easily characterized by labelling each $m$-rooted history of $t^{(n)}$ with the couple of integers $(n,m)$. Note that the
% labelling $(n,m)$ does not uniquely specify an $m$-rooted history of $t^{(n)}$.

% Let $H_{n,m}=H_{n,m}(t)$ be the set of $m$-rooted histories of $t^{(n)}$. Thus, $H_n = \bigcup_{m=1}^{\infty} H_{n,m}$ is the set of histories of $t^{(n)}$
% and $h_n = |H_n|$.
% We define the operator $\Omega$ as a function \footnote{$\mathcal{P}(X) = \{ x : x \subseteq X  \}$ denotes the power set of $X$.} $$\Omega: H_{n,m}
% \rightarrow \mathcal{P}(H_{n+1}).$$

\smallskip

\noindent \textbf{Definition.} Let $\mathcal{P}(X) = \{ x : x \subseteq X  \}$ denote the power set of set $X$, and fix tree $t$. The operator $\Omega$ is a function
$$\Omega: H_{n}(t) \rightarrow \mathcal{P}(H_{n+1}(t)),$$
where for a given rooted history $h \in H_n(t)$, $\Omega (h)$ is the set of rooted histories $h^\prime \in H_{n+1}(t)$ for which the restriction of $h^\prime$ to $t^{(n+1)}$ excluding its most basal caterpillar branch coincides with the rooted history $h$ of $t^{(n)}$.

%%%%%%%%%%%%%%%%%%%%%%%%%%%%%%%%%%%%%%%%%%%%%%%%%%%%%%%%%
%%%%%%%%%%%%%%%%%%%%%%%%%%%%%%%%%%%%%%%%%%%%%%%%%%%%%%%%%
%%%%%%%%%%%%%%%%%%%%%%%%%%%%%%%%%%%%%%%%%%%%%%%%%%%%%%%%%
\begin{figure}[tbh]
\begin{center}
\includegraphics*[scale=.72,trim=0 0 0 0]{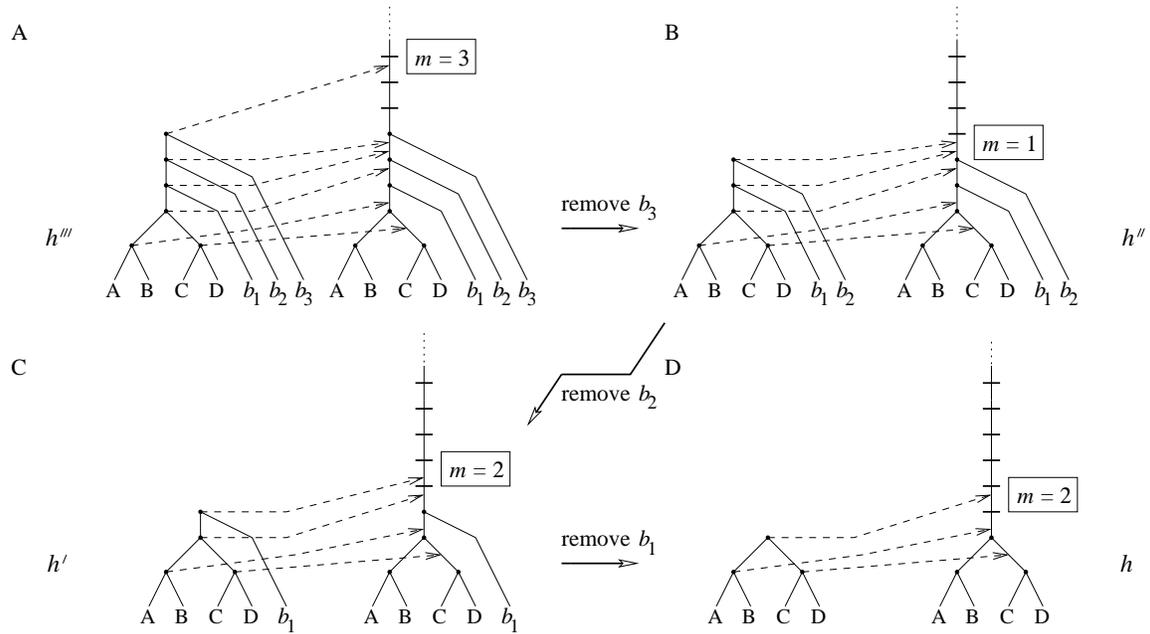}
\end{center}
\vspace{-.7cm}
\caption{{\small The relationships among rooted histories for sequential members of caterpillar-like families. For a rooted history $h^{\prime \prime \prime}$ of $t^{(3)}$, with $t=((A,B),(C,D))$, the figure sequentially removes caterpillar branches. By definition, a rooted history $h^\prime$ of $t^{(n+1)}$ belongs to the set $\Omega (h)$ if, by removing the most basal caterpillar branch $b_{n+1}$ in $t^{(n+1)}$, we recover the rooted history $h$ of $t^{(n)}$. Note that when we remove the basal caterpillar branch $b_{n+1}$ from $t^{(n+1)}$, the root of $t^{(n+1)}$---to which the branch $b_{n+1}$ is attached---becomes the boundary between the first and second components of the root-branch of $t^{(n)}$, and is depicted as a horizontal segment. (A) $h^{\prime \prime \prime} \in \Omega(h^{\prime \prime})$. (B) $h^{\prime \prime} \in \Omega(h^{ \prime})$. (C) $h^{\prime} \in \Omega(h)$. (D) $h$. For each rooted history, the value of the parameter $m$, representing the component of the root-branch that receives the image of the root, is shown.
}}\label{figIndietro5}
\end{figure}
%%%%%%%%%%%%%%%%%%%%%%%%%%%%%%%%%%%%%%%%%%%%%%%%%%%%%%%%%
%%%%%%%%%%%%%%%%%%%%%%%%%%%%%%%%%%%%%%%%%%%%%%%%%%%%%%%%%
%%%%%%%%%%%%%%%%%%%%%%%%%%%%%%%%%%%%%%%%%%%%%%%%%%%%%%%%%

\smallskip

Denote by $b_1, b_2, \ldots, b_{n+1}$ the caterpillar branches in $t^{(n+1)}$, from the least basal $b_1$ to the most basal $b_{n+1}$ (Fig.~\ref{figIndietro5}). Upon removal of the most basal caterpillar branch $b_{n+1}$ from $t^{(n+1)}$, the root of $t^{(n+1)}$---to which branch $b_{n+1}$ is attached---is replaced by a demarcation between the first and second components of the root-branch of $t^{(n)}$. For instance, in Fig.~\ref{figIndietro5}A, starting from tree $t=((A,B),(C,D))$, we consider $h^{\prime \prime \prime}$, a $3$-rooted history of $t^{(3)}$. By removing the most basal caterpillar branch $b_3$ of $t^{(3)}$, we reduce to the 1-rooted history $h^{\prime \prime}$ of $t^{(2)}$ (Fig.~\ref{figIndietro5}B). Next, by removing the caterpillar branch $b_2$ of $t^{(2)}$, we reduce to the 2-rooted history $h^\prime$ of $t^{(1)}$ (Fig.~\ref{figIndietro5}C). By removing the remaining caterpillar branch $b_1$ from $t^{(1)}$, we reduce to the 2-rooted history $h$ of $t=t^{(0)}$ (Fig.~\ref{figIndietro5}D). Therefore, by the definition of $\Omega$, we have $h^\prime \in \Omega(h), h^{\prime \prime} \in \Omega(h^\prime)$, and $h^{\prime \prime \prime} \in \Omega(h^{\prime \prime})$.

By definition, $\Omega$ has the property that for each rooted history $h^\prime \in H_{n+1}(t)$, with $n \geq 0$, there exists exactly one rooted history $h \in H_n(t)$ such that $h^\prime \in \Omega(h)$. In other words, for each $n\geq 0$, the set of rooted histories $H_{n+1}(t)$ can be partitioned as a disjoint union,
\begin{equation}\label{sqcup}
H_{n+1}(t) = \bigsqcup_{h \in H_n(t)} \Omega(h).
\end{equation}
The set $H_{n+1}(t)$ is therefore generated without double occurrences of any rooted history by applying $\Omega$ to the rooted histories in $H_n(t)$. It follows immediately that in performing $n$ iterations of $\Omega$ to obtain $\Omega[\ldots [\Omega[\Omega(H_0)]] \ldots]$ from the set $H_0$ of rooted histories of $t^{(0)}$, all the rooted histories of $t^{(n)}$ are generated exactly once.

%%%%%%%%%%%%%%%%%%%%%%%%%%%%%%%%%%%%%%%%%%%%%%%%%%%%%%%%%
%%%%%%%%%%%%%%%%%%%%%%%%%%%%%%%%%%%%%%%%%%%%%%%%%%%%%%%%%
%%%%%%%%%%%%%%%%%%%%%%%%%%%%%%%%%%%%%%%%%%%%%%%%%%%%%%%%%

\subsubsection{Labels for rooted histories}
\label{secFoglietta}

The operator $\Omega$, starting from the rooted histories of $t^{(n)}$, generates the rooted histories of $t^{(n+1)}$. In this section,
% we switch from counting rooted histories to counting multisets of labels. To this goal,
we introduce a labeling scheme, giving each $m$-rooted history $h$ of $t^{(n)}$ a label $L(h)=(n,m)$.
% For instance, looking at Fig.~\ref{albhist}A, the resulting label for the $3$-rooted history of $t^{(0)}$ termed $h$ is $(0,3)$, whereas the $1$-rooted
% history of $t^{(1)}$ termed $h$ and depicted in Fig.~\ref{albhist}B is labeled $(1,1)$.
% Note that,
We then describe how $\Omega$ acts on the labels of the rooted histories, characterizing the set of labels $L[\Omega(h)] = \{ L(h^\prime) : h^\prime \in \Omega(h) \}$.
% Among the rooted histories in $H_n = \bigcup_{m=1}^{\infty} H_{n,m}$ we are interested in those belonging to $H_{n,1}$ and, in particular, in counting the
% cardinality $|H_{n,1}| = h_n$ (\ref{accauno}), for every $n$.
Our goal is to represent each set $H_{n}$ of rooted histories of $t^{(n)}$ by the multiset of its labels, reducing the enumeration of $|H_{n,m}|$ to the problem of counting certain ordered pairs $(n,m)$ iteratively generated by simple rules that reflect how the rooted histories in $H_{n+1}$ are generated according to rule $\Omega$ from the rooted histories in $H_{n}$ by eq.~(\ref{sqcup}).

In our labeling scheme, each rooted history $h \in H_n(t)$ that maps the root of $t^{(n)}$ onto the $m$th component of the root-branch of $t^{(n)}$ receives label $L(h) = (n,m)$. The enumeration of $h_n=|H_{n,1}|$ then reduces to the enumeration of those rooted histories labeled by $(n,1)$.

Note that a label $(n,m)$ does not uniquely specify an $m$-rooted history of $t^{(n)}$: a tree $t^{(n)}$ has in general many $m$-rooted histories, each  receiving the label $(n,m)$. In other words, if $h, \overline{h} \in H_n(t)$ and $L(h) = L(\overline{h})$, then $h$ and $\overline{h}$ are not necessarily the same rooted history of $t^{(n)}$. We will, however, consider for $n \geq 0$ \emph{multisets} of labels in which we find a copy of the label $(n,m)$ for each $m$-rooted history of $t^{(n)}$.

\begin{figure}
\begin{center}
\includegraphics*[scale=.68,trim=0 0 0 0]{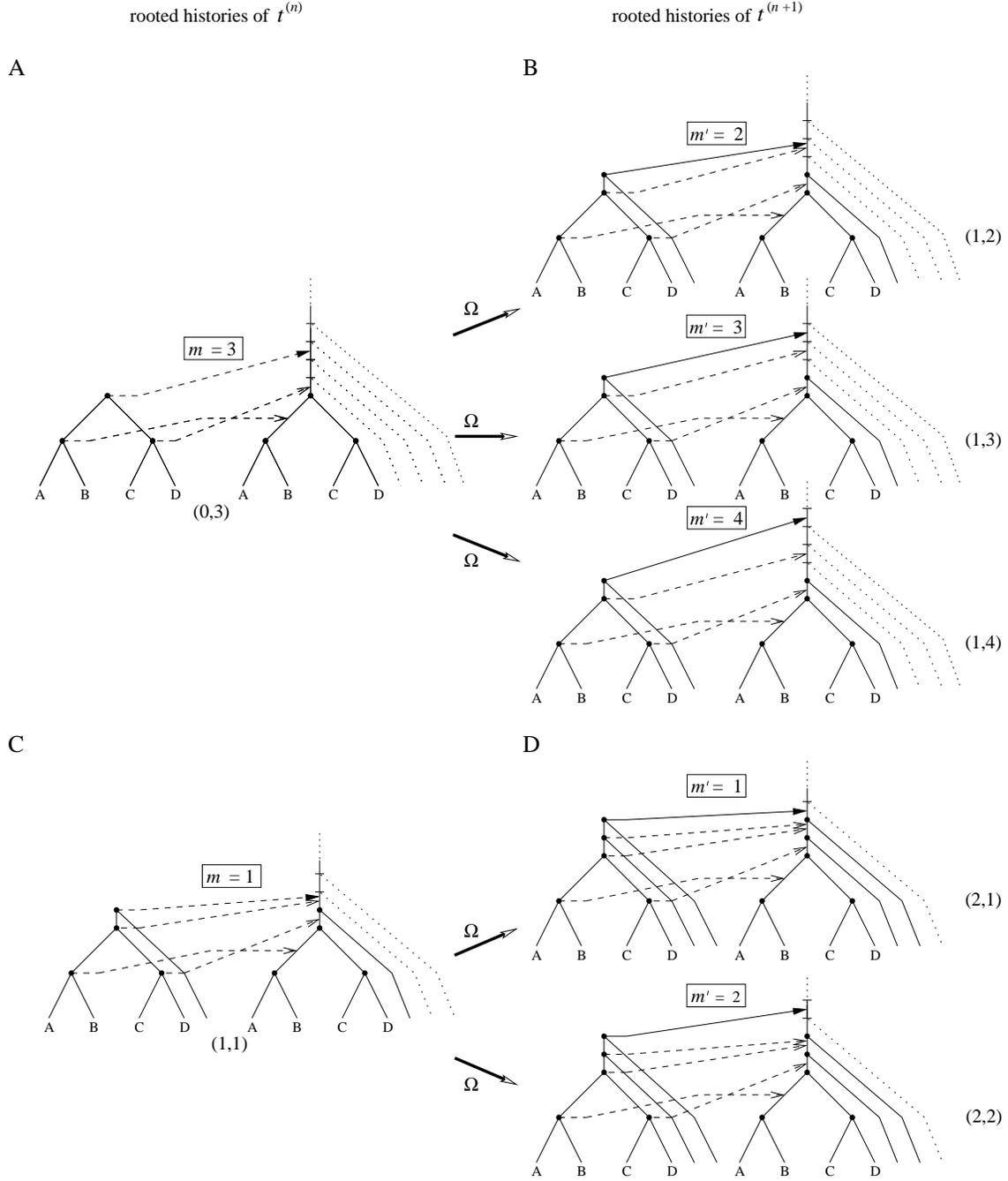}
\end{center}
\vspace{-.7cm}
\caption{{\small Generation of rooted histories of $t^{(n+1)}$ from rooted histories of $t^{(n)}$, as given by rule $\Omega$ applied to seed tree $t=((A,B),(C,D))$. To obtain rooted histories of $t^{(n+1)}$ (right) from rooted histories of $t^{(n)}$ (left), we choose the component $m^\prime$ of the root-branch of $t^{(n+1)}$ onto which the root of $t^{(n+1)}$ is mapped (solid arrows). The smallest among infinitely many possible choices are depicted. For all nodes of $t^{(n+1)}$ except the root, the rooted history generated for $t^{(n+1)}$ coincides with the generating rooted history of $t^{(n)}$ (dashed arrows). (A) A case with $m \geq 2$. A 2-rooted history $h$ of $t^{(0)}$, labeled $(0,3)$, is shown. (B) $\Omega(h)$ for $h$ in (A). 2-, 3-, and 4-rooted histories of $t^{(1)}$ belonging to $\Omega(h)$ are shown and are labeled $(1,2)$, $(1,3)$, and $(1,4)$, respectively. Because $m \geq 2$, $m^\prime \geq m-1$ as in eq.~(\ref{omeg}). (C) A case with $m=1$. A 1-rooted history $h$ of $t^{(1)}$, labeled $(1,1)$, is shown. (D) $\Omega(h)$ for $h$ in (C). 1- and 2-rooted histories of $t^{(2)}$ belonging to $\Omega(h)$ are shown and are labeled $(2,1)$ and $(2,2)$, respectively. Because $m=1$, $m^\prime \geq m$.}}
\label{figAlbhist6}
\end{figure}

To characterize how the operator $\Omega$ acts on the labels for rooted histories, consider an $m$-rooted history $h \in H_n(t)$, so that $h$ maps the root of $t^{(n)}$ onto the $m$th component of the root-branch of $t^{(n)}$. This history is labeled $L(h)=(n,m)$. For instance, taking the seed tree $t=((A,B),(C,D))$, the history $h$ of $t=t^{(0)}$ depicted in Figure \ref{figAlbhist6}A is labeled $L(h)=(0,3)$, whereas the history $h$ of $t^{(1)}$ in Figure \ref{figAlbhist6}C is labeled $L(h)=(1,1)$.

%%%%%%%%%%%%%%%%%%%%%%%%%%%%%%%%%%%%%%%%%%%%%%%%%%%%%%%%%
%%%%%%%%%%%%%%%%%%%%%%%%%%%%%%%%%%%%%%%%%%%%%%%%%%%%%%%%%
%%%%%%%%%%%%%%%%%%%%%%%%%%%%%%%%%%%%%%%%%%%%%%%%%%%%%%%%%
%%%%%%%%%%%%%%%%%%%%%%%%%%%%%%%%%%%%%%%%%%%%%%%%%%%%%%%%%
%%%%%%%%%%%%%%%%%%%%%%%%%%%%%%%%%%%%%%%%%%%%%%%%%%%%%%%%%
%%%%%%%%%%%%%%%%%%%%%%%%%%%%%%%%%%%%%%%%%%%%%%%%%%%%%%%%%

By applying $\Omega$ to a history $h$ of $t^{(n)}$ with $L(h)=(n,m)$, we produce a set of rooted histories $\Omega(h) \subseteq H_{n+1}(t)$. The set of labels for $\Omega(h)$,
$$ L[\Omega(h)]  = \{L(h^\prime) : h^\prime \in \Omega(h)  \},$$
is determined according to the rule:
\begin{equation}\label{omeg}
L[\Omega(h)] = \left\{
  \begin{array}{l l}
   \{ (n+1,m^\prime) :  m^\prime \geq m   \} & \text{if $m=1$}  \\
   \{ (n+1,m^\prime) :  m^\prime \geq m-1 \} & \text{if $m\geq 2$,}
  \end{array} \right.
\end{equation}
where $m^\prime$ denotes the value of the parameter $m$---the component of the root-branch of $t^{(n+1)}$ to which the root is mapped---for the rooted histories $h^\prime \in \Omega(h)$ of $t^{(n+1)}$.

The rule in eq.~(\ref{omeg}) distinguishes between two cases depending on whether the value of the parameter $m=m(h)$ of the generating rooted history $h$ is equal to or exceeds 1. In both cases, the set $L[\Omega(h)]$ contains infinitely many labels, each with its first component equal to $n+1$, as the labels refer to rooted histories of $t^{(n+1)}$. The value of the second component $m^\prime$ ranges in $[m-1,\infty)$ if $m\geq 2$, and in $[1,\infty)$ if $m=1$.

Recall that according to the definition of $\Omega$, from an $m$-rooted history $h$ of $t^{(n)}$ (Fig.~\ref{figAlbhist6}A and \ref{figAlbhist6}C), we generate an $m^\prime$-rooted history $h^\prime \in \Omega(h)$ of $t^{(n+1)}$ (Fig.~\ref{figAlbhist6}B and \ref{figAlbhist6}D) by (i) choosing the component $m^\prime$ of the root-branch of $t^{(n+1)}$ onto which $h^\prime$ maps the root of $t^{(n+1)}$, and (ii) letting $h^\prime$ coincide with $h$ on all nodes of $t^{(n+1)}$ except the root. The rooted history $h^\prime$ coincides with $h$ once we remove the most basal caterpillar branch of $t^{(n+1)}$.

Figure \ref{figAlbhist6} illustrates both cases of eq.~(\ref{omeg}). In step (i), infinitely many choices of $m^\prime$ are possible, because the root-branch of $t^{(n+1)}$ is divided into infinitely many parts. The most basal caterpillar branch in $t^{(n+1)}$ is attached at the border between the first and second components of the root-branch of $t^{(n)}$. Thus, the addition of the $(n+1)$st caterpillar branch eliminates a component of the root-branch, so that if the starting rooted history $h$ has $m \geq 2$ (Fig.~\ref{figAlbhist6}A), then the root of $t^{(n)}$ maps to component $m-1$ of the root-branch of $t^{(n+1)}$. The root of $t^{(n+1)}$ can map to this same branch, or to any branch $m^\prime$ with $m^\prime \geq m-1$. For instance, in Figure \ref{figAlbhist6}B, one of the rooted histories $h^\prime$ generated by a rooted history $h$ with $m=3$ has $m^\prime = m-1 = 2$.

If $h$ has $m=1$, however, then production of $h^\prime$ is slightly different (Fig.~\ref{figAlbhist6}C). By definition, the parameter $m$ for a rooted history cannot be smaller than $1$. The value $m^\prime = m-1$ is not permitted in this case, and $m^\prime$ remains greater than or equal to $m=1$ (Fig.~\ref{figAlbhist6}D).

% To better understand the correspondence between rooted histories and labels under the action of the operator $\Omega$, in Fig.~\ref{albhist} we have
% represented some applications of rule (\ref{omeg}) for two different rooted histories $h$, one in which $m=m(h) \geq 2$ (Fig.~\ref{albhist}A) and one in
% which $m=m(h)=1$ (Fig.~\ref{albhist}B). In (A), $h$ is a 3-rooted history of $t^{(0)}$ thus labelled $L(h)=(0,3)$. The set $\Omega(h)$ contains infinitely
% many rooted histories of $t^{(1)}$, one for each integer $m' \geq m-1 = 2$ (see case $m\geq 2$ in (\ref{omeg})). Three among the rooted histories in
% $\Omega(h)$ are depicted on the right, more precisely we depict the ones whose labels have second component respectively given by  $m'=2$, $m'=3$ and $m'=4$, % and which are thus labelled by $(1,2),(1,3)$ and $(1,4)$ respectively. In (B), $h$ is a 1-rooted history $L(h)=(1,1)$ of $t^{(1)}$. The application of
% $\Omega$ to $h$ produces infinitely many rooted histories of $t^{(2)}$, one for each integer $m' \geq 1$ (see case $m=1$ in (\ref{omeg})). Two among the
% rooted histories in $\Omega(h)$ are shown on the right: those whose labels have $m'=1$ and $m'=2$, and which are thus labelled by $(2,1)$ and $(2,2)$
% respectively.

%%%%%%%%%%%%%%%%%%%%%%%%%%%%%%%%%%%%%%%%%%%%%%%%%%%%%%%%%
%%%%%%%%%%%%%%%%%%%%%%%%%%%%%%%%%%%%%%%%%%%%%%%%%%%%%%%%%
%%%%%%%%%%%%%%%%%%%%%%%%%%%%%%%%%%%%%%%%%%%%%%%%%%%%%%%%%

\subsubsection{From counting rooted histories to counting their labels}
\label{secAsinello}

The labeling scheme in Section \ref{secFoglietta} encodes the application of the operator $\Omega$ to the rooted histories of $t^{(n)}$. Now that we have described the set of labels $L[\Omega(h)]$ arising from the label $L(h)$ according to the rule in eq.~(\ref{omeg}), the problem of counting a set of rooted histories becomes a problem of counting the set of the associated labels along with their multiplicities---or the \emph{multiset} of the labels.

\begin{figure}
\begin{center}
\includegraphics*[scale=.94,trim=0 0 0 0]{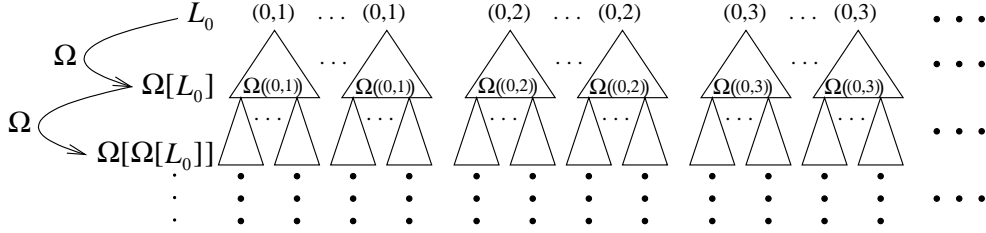}
\end{center}
\vspace{-.7cm}
\caption{{\small Iterative application of a rule for generating the multiset of the labels of the rooted histories of a tree $t^{(n)}$. The iterative procedure starts with the multiset $L_0$ that contains those labels of the form $\{(0,m) : m \geq 1 \}$ associated with the rooted histories of a seed tree $t = t^{(0)}$. In the first step of the iteration, we apply $\Omega$ (eq.~(\ref{omeg2})) to each label of $L_0$. In the second step, we apply $\Omega$ to each label resulting from the first step, and so on. The number of $m$-rooted histories of $t^{(n)}$ corresponds to the number of labels $(n,m)$, considered with their multiplicity, generated after the $n$th step of the iteration.
}} \label{figGenerazioni7} % NOTE: I CHANGED omeg3 TO omeg2 IN THIS CAPTION
\end{figure}

For $n \geq 0$ and $m \geq 1$, we use $\Omega\big( (n,m) \big)$ to denote, with an abuse of notation, the set of labels $L[\Omega(h)]$ when $L(h)=(n,m)$. Recalling that iterative application of $\Omega$ to the rooted histories $H_0$ of tree $t^{0)}$ generates the rooted histories $H_n$ of $t^{(n)}$, the enumeration of $|H_{n,m}|$ for tree $t=t^{(0)}$ becomes a problem of counting those labels of the form $(n,m)$ that are generated when we iteratively apply the operator $\Omega$ as $\Omega[ \ldots [\Omega[\Omega(L_0)]] \ldots]$ starting from the multiset of labels $L_0= \{L(h) : h \in H_0(t) \}$ (Fig.~\ref{figGenerazioni7}).

%%%%%%%%%%%%%%%%%%%%%%%%%%%%%%%%%%%%%%%%%%%%%%%%%%%%%%%%%
%%%%%%%%%%%%%%%%%%%%%%%%%%%%%%%%%%%%%%%%%%%%%%%%%%%%%%%%%
%%%%%%%%%%%%%%%%%%%%%%%%%%%%%%%%%%%%%%%%%%%%%%%%%%%%%%%%%
%%%%%%%%%%%%%%%%%%%%%%%%%%%%%%%%%%%%%%%%%%%%%%%%%%%%%%%%%
%%%%%%%%%%%%%%%%%%%%%%%%%%%%%%%%%%%%%%%%%%%%%%%%%%%%%%%%%
%%%%%%%%%%%%%%%%%%%%%%%%%%%%%%%%%%%%%%%%%%%%%%%%%%%%%%%%%

% Our goal is indeed to represent each set $H_{n}$ of rooted histories of $t^{(n)}$ by the multiset of its labels reducing the enumeration of $|H_{n,m}|$ to
% the problem of counting certain ordered pairs of positive integers $(n,m)$ iteratively generated by simple rules (\ref{omeg2}) that reflect how the rooted
% histories in $H_{n+1}$ are generated according to rule $\Omega$ starting from the rooted histories in $H_{n}$ (\ref{sqcup}).

Eq.~(\ref{omeg}) characterizes the set of labels $L[\Omega(h)]$ of the rooted histories in $\Omega(h)$ in terms of the label $L(h)$ of rooted history $h$. If $L(h)=(n,m)$, then $\Omega\big( (n,m) \big)$ denotes the set of labels $L[\Omega(h)]$. Thus, converting the notation from histories to labels, eq.~(\ref{omeg}) becomes
\begin{equation}\label{omeg2}
\Omega\big( (n,m) \big) = \left\{
  \begin{array}{l l}
   \{ (n+1,m') :  m' \geq m  \} &  \text{if $m=1$}\\
   \{ (n+1,m') :  m' \geq m-1\} & \text{if $m\geq 2$.}
  \end{array} \right.
\end{equation}
For the seed tree $t$, we count $h_{n,m} = |H_{n,m}|$ by evaluating the number of occurrences of the ordered pair $(n,m)$ in the multiset $L_n$ defined as
\begin{equation}\label{babbino}
L_n = L[H_{n}(t)] = \{L(h) : h \in H_n(t) \}.
\end{equation}
In symbols, we have
\begin{equation}\label{ricordi}
h_{n,m} =  |\{\ell \in L_n : \ell = (n,m) \}|.
\end{equation}

By eq.~(\ref{sqcup}), each multiset $L_n$ is generated iteratively (Fig.~\ref{figGenerazioni7}). We start with the multiset of labels
\begin{equation}\label{zero}
L_0 = \{L(h) : h \in H_0(t) \}.
\end{equation}
For each $n \geq 0$, the multiset $L_{n+1}$ is obtained as the union
\begin{equation}\label{dopo}
L_{n+1} = \biguplus_{(n,m) \in L_n}  \Omega\big( (n,m)  \big),
\end{equation}
where the symbol $\biguplus$ denotes the union operator for multisets. Thus, in $M=M_1 \biguplus M_2$, if an element $x$ appears $n_1$ times in $M_1$ and $n_2$ times in $M_2$, then it appears $n_1+n_2$ times in $M$. Eq.~(\ref{dopo}) provides an iterative generation of the labels for the rooted histories of $H_{n+1}(t)$ from the labels of the rooted histories of $H_n(t)$, retaining information about the multiplicity of occurrences of each label.

% Once a rooted history labeled $(\overline{n},\overline{m})$ has been fixed, by iterating $\Omega$ we can build a \emph{generating tree} \cite{bosc, eco} of
% rooted histories descending from that one labelled $(\overline{n},\overline{m})$. In this tree, each node is a label of the form $(n,m)$ whose direct
% descendants are the infinitely many labels belonging to the set $\Omega(n,m)$ as defined in (\ref{omeg}). Each descendant label generates itself new
% nodes/labels in the tree by applying again $\Omega$ and so on.

% This is depicted in Fig.~\ref{albhist}.
% Note that to generate a rooted history of $t^{(n+1)}$ from one of $t^{(n)}$, only the image $m$ of the root in $t^{(n+1)}$ must be determined. Observe that
% $m$ can increase without bounds. If the generating history is labelled $m=1$, then in the newly generated history $m$ cannot decrease. Otherwise, from $m\geq % 2$ we can also create an history with $m=m-1$.

%%%%%%%%%%%%%%%%%%%%%%%%%%%%%%%%%%%%%%%%%%%%%%%%%%%%%%%%%
%%%%%%%%%%%%%%%%%%%%%%%%%%%%%%%%%%%%%%%%%%%%%%%%%%%%%%%%%
%%%%%%%%%%%%%%%%%%%%%%%%%%%%%%%%%%%%%%%%%%%%%%%%%%%%%%%%%

\subsection{Counting rooted histories with generating functions}
\label{secGenfun}

We have now obtained eq.~(\ref{ricordi}), which gives an equivalence between the number of $m$-rooted histories of $t^{(n)}$ and the number of labels $(n,m)$ in the multiset $L_n$, and eqs.~(\ref{zero}) and (\ref{dopo}), which give through $\Omega$ (eq.~(\ref{omeg2})) an iterative procedure that generates the family of multisets $(L_n)_{n \geq 0}$. In this section, we translate the iterative procedure into algebraic terms, determining the generating function associated with the integer sequence $(h_n)_{n\geq 0}$.

First, in Section \ref{secComincio}, we characterize a generating function $g(y)$ for the sequence $(h_{0,m})_{m \geq 1}$. Next, in Section \ref{secEquazzz}, we deduce an equation satisfied by the bivariate generating function $F(y,z)$ for $(h_{n,m})_{n \geq 0, m \geq 1}$. In Section \ref{secSpecchio}, we solve the equation, obtaining the desired generating function $f(z)$ for the sequence $(h_{n,1})_{n \geq 0}$. This generating function can be written in turn as a function of $g(y)$.

\subsubsection{Generating function for the sequence $(h_{0,m})_{m\geq 1}$}
\label{secComincio}

In this section, we characterize the generating function $g(y)$ that counts for a given seed tree $t$ the labels in the multiset $L_0$ describing the labels of the rooted histories of $t$.

Fix the seed tree $t$. Recalling the equivalence in eq.~(\ref{ricordi}), define the generating function
\begin{equation}\label{g}
g(y) =  \sum_{(0,m) \in L_0} y^m = \sum_{m=1}^\infty h_{0,m} y^m,
\end{equation}
the $m$th coefficient of whose power series expansion provides the number $h_{0,m}$ of labels $(0,m)$ appearing in $L_0$. By Proposition~\ref{peperoncino}, $h_{0,m}$ can be expressed as a polynomial in the variable $m$ and can thus be decomposed as a finite linear combination of terms of the form $m^k$, where $k$ is a non-negative integer. That is, for a certain finite set of non-negative integers with largest element $K$,
\begin{equation}\label{gerno}
h_{0,m} = \sum_{k=0}^K w_k  m^k,
\end{equation}
where the $w_k$ are constants.

We introduce generating functions $g_{m^k}$, one for each $k$ from 0 to $K$, in which the $m$th coefficient is $m^k$:
\begin{equation}\label{monomi}
g_{m^k}(y) = \sum_{m=1}^\infty m^k y^m.
\end{equation}
Because $K$ is finite, the desired generating function $g(y)$ can be written as a finite linear combination of this new collection of generating functions $g_{m^0}(y), g_{m^1}(y), \ldots, g_{m^K}(y)$. More precisely, by substituting in eq.~(\ref{g}) the polynomial in eq.~(\ref{gerno}) and switching the order of summation, we obtain
\begin{equation}\label{minnetta}
g(y) = \sum_{k=0}^K w_k  g_{m^k}(y).
\end{equation}
We now state a lemma that characterizes the generating functions $g_{m^k}(y)$.
\begin{lemm}
For each non-negative integer $k$ from 0 to $K$, the generating function $g_{m^k}(y)$ in eq.~(\ref{monomi}) is rational with denominator $(1-y)^{k+1}$. That is, $g_{m^k}(y)$ has the form
\begin{equation*} %\label{bubbu}
g_{m^k}(y)= \frac{\text{P}(y)}{(1-y)^{k+1}},
\end{equation*}
where $P(y)$ is a polynomial in $y$.
\end{lemm}
\noindent {\bf Proof.}  We proceed by induction on $k$. If $k=0$, then by eq.~(\ref{monomi}), $g_{m^0}(y) = 1/(1-y)-1 = y/(1-y)$. Assume the inductive hypothesis for $g_{m^k}(y)$. Applying eq.~(\ref{monomi}) to $g_{m^{k+1}}(y)$, we can recover $g_{m^{k+1}}(y)$ as
\begin{equation}\label{ggg}
g_{m^{k+1}}(y) = y \frac{\partial g_{m^k}(y)}{\partial y},
\end{equation}
which by the quotient rule for derivatives is a rational function with denominator $(1-y)^{k+2}$. $\Box$

\medskip

The proof of the lemma gives a recursive procedure in eq.~(\ref{ggg}) to compute the functions $g_{m^k}(y)$. By eq.~(\ref{minnetta}), we immediately obtain from the lemma a result about the generating function $g(y)$.

\begin{prop}\label{pinco}
The generating function $g(y)$ whose $m$th coefficient $[y^m]g(y)$ is the number of $m$-rooted histories $h_{0,m}$ of a seed tree $t$ can be written as a finite linear combination
\begin{equation}\label{combin}
g(y) = \sum_{j=1}^J q_j \frac{y^{a_j}}{(1-y)^{b}},
\end{equation}
where $b\geq 1$ and $J \geq 1$ are positive integers, each $a_j$ is a non-negative integer, and the $q_j$ are constants.
\end{prop}
% \emph{Proof.} When $k=0$, we have $g_{m^0}(z) = 1/(1-z)-1$. If $k=i+1$, then
% \begin{equation}\label{ggg}
% g_{m^{i+1}}(z) = z \frac{\partial g_{m^i}}{\partial z},
% \end{equation}
% which has denominator $(1-z)^{i+2}$. Finally, note that $g$ can be written as finite linear combination of the functions $g_{m^i}$ $(0 \leq i \leq k)$, and
% therefore has denominator $(1-z)^{k+1}$. \,\,\, $\Box$

\medskip

As an example, we show how the procedure in Proposition~\ref{pinco} can be applied to determine the generating function $g(y)$ for $t=((A,B),(C,D))$, the same example seed tree for which we computed the polynomial $h_{0,m}$ via Proposition~\ref{peperoncino}. Recall from Section \ref{secRelations} that $h_{0,m} = m^2+2m+1$. To obtain the generating function $g(y)$ that has coefficients $[y^m]g(y)= m^2+2m+1,$ we sum generating functions for the monomials $m^2$, $2m$, and 1. We already know $g_{m^0}(y)$, and by applying
eq.~(\ref{ggg}), we have
\begin{eqnarray}
g_{m^0}(y) & = & \frac{y}{1-y} \nonumber \\
g_{m^1}(y) & = & y \frac{\partial g_{m^0}(y)}{\partial y} = \frac{y}{(1-y)^2} \nonumber \\
g_{m^2}(y) & = & y \frac{\partial g_{m^1}(y)}{\partial y} = \frac{y(y+1)}{(1-y)^3}. \nonumber
\end{eqnarray}
Thus,
\begin{equation}\label{gio}
g(y) = g_{m^0}(y) + 2 g_{m^1}(y) + g_{m^2}(y) = \frac{y^3 - 3y^2 + 4y}{(1-y)^3}.
\end{equation}
In eq.~(\ref{gio}), $g(y)$ is written as in eq.~(\ref{combin}), taking $b=3$, $J=3$, $(a_1,a_2,a_3) = (1,2,3)$, and $(q_1,q_2,q_3) = (4,-3,1)$.

%%%%%%%%%%%%%%%%%%%%%%%%%%%%%%%%%%%%%%%%%%%%%%%%%%%%%%%%%
%%%%%%%%%%%%%%%%%%%%%%%%%%%%%%%%%%%%%%%%%%%%%%%%%%%%%%%%%
%%%%%%%%%%%%%%%%%%%%%%%%%%%%%%%%%%%%%%%%%%%%%%%%%%%%%%%%%

\subsubsection{Bivariate generating function for the integers $(h_{n,m})_{n \geq 0, m \geq 1}$}
\label{secEquazzz}

Given $t$, the polynomial nature of $h_{0,m}$ in $m$ enabled us to obtain a generating function for $h_{0,m}$. We now use the iterative procedure in eq.~(\ref{dopo}) to determine an equation that characterizes the bivariate generating function with coefficients $h_{n,m}$. We represent each label of the form $(n,m)$ by a symbolic algebraic expression in the variables $y$ and $z$, so that $(n,m)$ is replaced by $z^n y^m$. Let $L = \cup_{n=0}^\infty L_n$ be the multiset of all $m$-rooted histories for all trees $t^{(n)}$. Considering $y$ and $z$ as complex variables in two sufficiently small neighborhoods of 0, we aim to characterize the bivariate function $F(y,z)$ that admits the expansion
%The bivariate generating function
$$F(y,z) = \sum_{(n,m) \in L} z^n y^m,$$
where the sum is over all labels in the multiset $L$ and thus has a term for each $m$-rooted history of each $t^{(n)}$. In particular, the function $F(y,z)$ is the bivariate generating function of the integers $h_{n,m}$, and its Taylor expansion can be written as
\begin{equation}\label{f}
F(y,z) = \sum_{m=1}^\infty \sum_{n=0}^\infty h_{n,m} \, z^n y^m,
%= (h_{0,1}z^0 + h_{1,1}z^1 + \ldots)y^1 + (h_{0,2}z^0 + h_{1,2}z^1 + \ldots)y^2 + \ldots \, .
\end{equation}
where the coefficients $h_{n,m}$ appear explicitly.

By differentiating $F(y,z)$ with respect to $y$ and then taking $y=0$, we obtain
\begin{equation}
\label{eqDeriv}
\frac{\partial F}{\partial y}(0,z) = \sum_{n=0}^\infty h_{n,1}z^n. % h_{0,1}z^0 + h_{1,1}z^1 + \ldots
\end{equation}
Thus, for each $n \geq 0$, we have
\begin{equation*} %\label{oss}
h_n = h_{n,1} = [z^n] \bigg( \frac{\partial F}{\partial y}(0,z) \bigg).
\end{equation*}

% Assuming from now on that the complex variables $z$ and $y$ are close to $0$, $z \approx 0$ and $y \approx 0$, we
By representing each label of the form $(n,m)$ by the symbolic expression $z^n y^m$ and assuming the complex variables $y$ and $z$ are sufficiently close to 0, the recursive generation in eq.~(\ref{dopo}) of the multisets of labels $L_0, L_1, L_2, \ldots$  determines an equation for $F(y,z)$, demonstrated in Appendix 1:

\begin{equation}\label{scarpa}
F(y,z) \left[ 1 - \frac{z}{y(1-y)}  \right] = g(y) - z \frac{\partial F}{\partial y}(0,z).
\end{equation}
Eq.~(\ref{scarpa}) holds if the complex variables $y$ and $z$ are in two sufficiently small neighborhoods of 0, and it characterizes the generating function $F(y,z)$.

% Note that indeed we can legitimately substitute $y=0$ in (\ref{scarpa}) given that the $y$ that appears in the denominator of the left-hand side simplifies
% within the power series expansion of $F(y,z)$ (see (\ref{f})).

% In particular, given that $g(y)$ has a non-removable singularity at $y=1$ (see Proposition~\ref{pinco}), equality (\ref{scarpa}) has no meaning for $y=1$.

%%%%%%%%%%%%%%%%%%%%%%%%%%%%%%%%%%%%%%%%%%%%%%%%%%%%%%%%%
%%%%%%%%%%%%%%%%%%%%%%%%%%%%%%%%%%%%%%%%%%%%%%%%%%%%%%%%%
%%%%%%%%%%%%%%%%%%%%%%%%%%%%%%%%%%%%%%%%%%%%%%%%%%%%%%%%%

\subsubsection{Generating function for the sequence $(h_{n,1})_{n\geq 0}$}
\label{secSpecchio}

We now have an equation satisfied by the bivariate generating function $F(y,z)$. Further, we have eq.~(\ref{eqDeriv}), which demonstrates that the desired generating function for the sequence $(h_n)_{n \geq 0}$ is obtained from $\frac{\partial F}{\partial y}(0,z)$. By applying the \emph{kernel method} \cite{BanderierEtAl02, Prodinger04}, we can determine the power series $\frac{\partial F}{\partial y}(0,z)$ from eq.~(\ref{scarpa}).

The idea of the method consists of coupling the two variables $(z,y)$ as $(z,y(z))$ in such a way that two conditions hold. First, (i) substituting $y=y(z)$ cancels the \emph{kernel} of the equation, that is, the factor $1-z/[y(1-y)]$ on the left-hand side of eq.~(\ref{scarpa}). Second, (ii) for $z$ near $0$, the value of $y(z)$ remains in a sufficiently small neighborhood of $y=0$, so that eq.~(\ref{scarpa}) still holds near $z=0$ after substituting $y=y(z)$. This condition is required, as the power series expansion in eq.~(\ref{f}) for $F(y,z)$ has been assumed to be valid in a neighborhood of $(y,z)=(0,0)$, and the derivation of eq.~(\ref{scarpa}) relies on the fact that $y$ and $z$ are sufficiently close to 0. If the two conditions hold, then
$$z \frac{\partial F}{\partial y}(0,z) = g(y(z)),$$
so that $g(y(z))$ must be a power series for $z=0$, because so must be $z \frac{\partial F}{\partial y}(0,z)$.

The required substitution couples $y$ and $z$ in such a way that $1-z/[y(1-y)]=0$, so that $y(z) = (1\pm \sqrt{1-4z})/2$. To determine whether to take the negative root $y_1(z)$ or the positive root $y_2(z)$, we note that if $z$ is near 0, then $y_1(z)$ approaches 0, so that $y_1(z)$ lies in a neighborhood of $y=0$ and $g(y_1(z))$ admits a power series expansion for $z$ near 0. For $y_2(z)$, however, if $z$ is near 0, then $y_2(z)$ approaches $1$, and thus, $g(y_2(z))$ is not a power series for $z$ near 0 due to the pole of the function $g(y)$ at $y=1$ (Proposition~\ref{pinco}).
%and $y_2(z)$ does not satisfy (ii)\fil{[[]]}.
The only solution satisfying both (i) and (ii) is consequently

% \textcolor{red}{Indeed, $g(y_2(z))$ is not a power series at $z=0$ but a \emph{Laurent} series; its expansion at $z=0$ starts with $1/z^{\alpha}$, for some
% $\alpha >0 $, due to the pole at $y=1$ for $g(y)$ (Proposition~\ref{pinco}).}

% We know by Proposition~\ref{pinco} that $g(y)$ has a pole at $y=1$ and thus $g(y_2(z))$ does not give a power series at $z=0$ (but a Laurent one). In other
% words, the neighborood $B_y(0)$ --- the domain of the variable $y$ for which (\ref{scarpa}) holds true --- is not big enough to contain $y_2(z)$ when $z$
% approaches $0$.

\begin{equation} \label{ker}
Y(z)=y_1(z) = \frac{1 - \sqrt{1-4z}}{2},
\end{equation}
which, with the generating function $C(z)$ of the Catalan numbers as in eq.~(\ref{funcat}), satisfies $Y(z) = z C(z)$. Substituting $y = Y(z)$ in eq.~(\ref{scarpa}), we have $\frac{\partial F}{\partial y}(0,z) = {g(Y(z))}/{z}$, yielding the following result.

\begin{prop}\label{pallino}
Fix tree $t$. Let $g(y)$ be the generating function associated with the polynomial $h_{0,m}$ (eq.~(\ref{g})). Let $Y(z)$ be as in eq.~(\ref{ker}). Then the generating function $f(z) = \sum_{n=0}^\infty h_n z^n$ is given by
\begin{equation}\label{zanza}
f(z) = \frac{\partial F}{\partial y}(0,z) = \frac{g(Y(z))}{z}  = \frac{g\big(\frac{1-\sqrt{1-4z}}{2}\big)}{z}.
\end{equation}
\end{prop}

The proposition thus determines the generating function $f(z)= g(Y(z))/z$ for the integer sequence describing the number of matching coalescent histories of the species trees in the caterpillar-like family $(t^{(n)})_{n\geq 0}$. The function $g$ depends on the seed tree $t$, whereas the function $Y(z)$ is fixed in eq.~(\ref{ker}) and does not depend on $t$.

As an example, recall that for $t = ((A,B),(C,D))$, in eq.~(\ref{gio}), we have computed the generating function $g$ for the number $h_{0,m}$ of $m$-rooted histories of $t=t^{(0)}$. By Proposition~\ref{pallino}, the generating function for the number $h_n$ of matching coalescent histories of $t^{(n)}$ is
$$f(z)  = \sum_{n=0}^\infty h_n z^n  =\frac{g\big( \frac{1-\sqrt{1-4z}}{2} \big)}{z} = \frac{4(1-\sqrt{1-4z})(3-z+\sqrt{1-4z})}{z (1+\sqrt{1-4z})^3}.$$
Taking the Taylor expansion of $f$, we obtain
\begin{equation}
\label{eqTaylor}
f(z)= 4 + 13 z + 42 z^2 + 138 z^3 + 462 z^4 + 1573 z^5 + 5434 z^6 + 19006 z^7 + 67184 z^8 + \ldots
\end{equation}
The coefficients $h_n$ accord with the enumeration of matching coalescent histories reported in Corollary 3.9 of \cite{Rosenberg07:jcb} and Table 3 of \cite{Rosenberg13:tcbb} for caterpillar-like families with seed tree $t=((A,B),(C,D))$, except that those results tabulated numbers of coalescent histories by the number of taxa, whereas here, we use the index of the caterpillar-like family. Thus, in this example, the coefficient of $z^n$ gives the number of matching coalescent histories for a tree with $n+4$ taxa, as $|t|=4$. Shifting the index in the formula from \cite{Rosenberg07:jcb, Rosenberg13:tcbb} to agree with our indexing scheme, we obtain $[(5(n+4)-12)/(4(n+4)-6)]c_{(n+4)-1} = [(5n+8)/(4n+10)]c_{n+3}$ for the number of matching coalescent histories of $t^{(n)}$. This formula gives precisely the coefficients in the Taylor expansion in eq.~(\ref{eqTaylor}).

%%%%%%%%%%%%%%%%%%%%%%%%%%%%%%%%%%%%%%%%%%%%%%%%%%%%%%%%%
%%%%%%%%%%%%%%%%%%%%%%%%%%%%%%%%%%%%%%%%%%%%%%%%%%%%%%%%%
%%%%%%%%%%%%%%%%%%%%%%%%%%%%%%%%%%%%%%%%%%%%%%%%%%%%%%%%%

\subsection{Asymptotic behavior of $h_n$}
\label{secSomaro}

From Proposition~\ref{pallino}, we have the generating function $f$ that counts the number of matching histories of $t^{(n)}$ for a given fixed seed tree $t$. Applying techniques of analytic combinatorics as introduced in Section~\ref{secMethods}, we can determine the asymptotic behavior of the coefficients of the generating function
\begin{equation}\label{titti}
\tilde{f}(z) = \sum_{n=1}^\infty h_{n-1}z^n   =  z f(z) = g(Y(z)),
\end{equation}
with $Y(z)$ as in eq.~(\ref{ker}). To simplify notation, we work with $\tilde{f}$ instead of $f$.

First, in Section \ref{secGeneralAsymptotic}, we obtain an asymptotic equivalence between $h_n$ and $\beta_t c_n$, where $\beta_t$ is a constant depending on the seed tree $t$, and the $c_n$ are the Catalan numbers (eq.~(\ref{catalano1})). Next, in Section \ref{secConnection}, we produce a general procedure to determine the constants $\beta_t$, employing this procedure to obtain the values of $\beta_t$ for all seed trees $t$ with $|t| \leq 9$. We demonstrate that our values of $\beta_t$ accord with constant multiples of the Catalan numbers previously obtained according to a different method \cite{Rosenberg13:tcbb} for seed trees with $|t| \leq 8$.

\subsubsection{A general asymptotic result}
\label{secGeneralAsymptotic}

Recall that given $t$, Proposition~\ref{pinco} gives a procedure to determine the rational function $g$ in eq.~(\ref{titti}). Writing $g$ as the finite linear combination in eq.~(\ref{combin}), the values of $b$, $J$, and the $(a_j)_{1 \leq j \leq J}$ and $(q_j)_{1 \leq j \leq J}$ can all be computed.

% Further details and references can be found in Chapters IV and VI of \cite{ancomb}.

% From Proposition~\ref{pinco}, $g(z)$ is a rational function with denominator of the form $(1-z)^{b}$ with $b \geq 1$. See for instance (\ref{gio}), which is % obtained for the case $t=(()())$. Clearly, $g(z)$ can be written as a finite linear combination
% \begin{equation}\label{combin}
% g(z) = \sum_{i=1}^J k_i \frac{z^{a_i}}{(1-z)^{b}}.
% \end{equation}

As noted in Section \ref{secMethods}, the expansion of $\tilde{f}$ at its dominant singularity characterizes the asymptotic behavior of the coefficients $h_{n-1}$. In Appendix 2, we obtain the expansion of $\tilde{f}$ at the dominant singularity $z=\frac{1}{4}$,

\begin{eqnarray}\label{asino}
\tilde{f}(z) & =    & \alpha_t + \beta_t \bigg(-\frac{\sqrt{1-4z}}{2}\bigg) \pm \mathcal{O}(1-4z)  \\\label{creek}
             & \sim & \alpha_t + \beta_t \bigg(-\frac{\sqrt{1-4z}}{2}\bigg),
\end{eqnarray}
with
\begin{eqnarray}\label{be1}
\alpha_t & = & \sum_{j=1}^J 2^{b-a_j}          q_j  \\
\label{be2}
\beta_t  & = & \sum_{j=1}^J 2^{b+1-a_j}(a_j+b) q_j.
\end{eqnarray}

Note that in eq.~(\ref{asino}), the seed tree affects only the constants $\alpha_t$ and $\beta_t$ computed in eqs.~(\ref{be1}) and (\ref{be2}) from $g$, as written in the linear combination in eq.~(\ref{combin}). Excluding the constant $\alpha_t$ that does not influence the asymptotic behavior of the coefficients, the main term of the expansion of $\tilde{f}(z)$ (eq.~(\ref{creek})) is the product of the constant $\beta_t$ and the generating function $-\sqrt{1-4z}/2$, whose $n$th coefficient is the Catalan number $c_{n-1}$ (eq.~(\ref{lucia1})).

Theorem VI.4 of \cite{FlajoletAndSedgewick09} indicates that under conditions satisfied by $\tilde{f}$, the asymptotic coefficients of a generating function as $n \rightarrow \infty$ are obtained from the expansion of the function at the dominant singularity; moreover, the error term in the asymptotic coefficients can be computed from the error term in the singular expansion. Applying the theorem to the expansion in eq.~(\ref{asino}), we obtain the asymptotic behavior of the coefficients $[z^n] \tilde{f}(z) = h_{n-1}$.
\begin{prop}\label{basta}
For any seed tree $t$, when $n \rightarrow \infty$, the number $h_n$ of matching coalescent histories for $t^{(n)}$ satisfies
\begin{equation}\label{tira}
h_{n-1} = [z^n] \tilde{f}(z) \sim \beta_t  [z^n] \bigg(- \frac{\sqrt{1-4z}}{2} \bigg) \pm \mathcal{O}\bigg( \frac{4^n}{n^2}  \bigg)
= \beta_t c_{n-1} \pm \mathcal{O}\bigg( \frac{4^n}{n^2}  \bigg),
\end{equation}
where $\beta_t$ is a constant that depends on $t$. The constant $\beta_t$ is computed in eq.~(\ref{be2}) once the function $g$, which is defined in eq.~(\ref{g}), has been written as the linear combination in eq.~(\ref{combin}).
\end{prop}

We immediately obtain the following corollary, corresponding to our initial claim in eq.~(\ref{equi}).

\begin{coro}\label{pluto}
For any seed tree $t$, there exists a constant $\beta_t > 0$ (eq.~(\ref{be2})) such that when $n \rightarrow \infty$,
\begin{equation}\label{ecci}
h_n \sim \beta_t c_n.
\end{equation}
\end{coro}

\noindent {\bf Proof.} The result follows from Proposition~\ref{basta} by noting that if $\beta_t > 0$, then
$$\lim_{n \rightarrow \infty} \frac{h_{n-1}}{\beta_t c_{n-1}} = 1 \pm \lim_{n \rightarrow \infty} \frac{\mathcal{O}(4^n/n^2)}{\beta_t c_{n-1}} = 1.$$

Note that we are claiming $\beta_t > 0$. From the definition of $\beta_t$ as the sum in eq.~(\ref{be2}), because the $q_j$ are permitted to be negative, it is not immediately clear that $\beta_t > 0$. Proposition \ref{basta} eliminates the possibility that $\beta_t$ is negative, as $h_{n-1}$ is necessarily positive. To show that $\beta_t \neq 0$, we note that by eq.~(\ref{tira}), $\beta_t = 0$ would give
\begin{equation}\label{sissa}
h_{n-1} = \mathcal{O}\bigg( \frac{4^n}{n^2} \bigg),
\end{equation}
so that $h_{n-1}/(4^n/n^2)$ would remain bounded by a constant as $n \rightarrow \infty$.

We now apply the lower bound $h_n \geq c_{n+1}$ from eq.~(\ref{lov}). By eq.~(\ref{lov}), we have
$$\frac{h_{n-1}}{{4^n}/{n^2}} \geq \frac{c_{n}}{{4^n}/{n^2}} = \frac{\sqrt{n}}{\sqrt{\pi}} \frac{c_{n}}{{4^{n}}/({n^{3/2} \sqrt{\pi}})}.$$
As $n \rightarrow \infty$, $\sqrt{n}/\sqrt{\pi}$ diverges to $\infty$, while $c_{n}/[4^{n}/(n^{3/2} \sqrt{\pi})]$ converges to 1 by eq.~(\ref{lucia3}). Therefore, the sequence $h_{n-1}/(4^n/n^2)$ must diverge and eq.~(\ref{sissa}) cannot hold. Thus, $\beta_t \neq 0$. $\Box$

\medskip

As an example of Corollary \ref{pluto}, consider $t=((A,B),(C,D))$. By decomposing the function $g$ expressed in eq.~(\ref{gio}) as in eq.~(\ref{combin}), we have already obtained the parameters $b$, $J$, $(a_j)_{1\leq j \leq J}$, and $(q_j)_{1 \leq j \leq J}$ in Section \ref{secComincio}. Therefore, computing $\beta_t$ as in eq.~(\ref{be2}), we obtain
$$\beta_t = 2^{1+3-1}(1+3)(4) + 2^{1+3-2}(2+3)(-3) + 2^{1+3-3}(3+3)(1)=80.$$
Eq.~(\ref{ecci}) then produces $h_{n} \sim 80 c_n$. Note that the limit $h_n \sim \frac{5}{4}c_{n+3}$ produced for this tree from $h_n = [(5n+8)/(4n+10)]c_{n+3}$ in Section \ref{secSpecchio} agrees with the limiting result $h_n \sim 80c_n$. Recalling eq.~(\ref{eugenio}),
$$\frac{h_n}{c_n} = \frac{5n+8}{4n+10}\frac{c_{n+3}}{c_n} \sim \frac{5}{4} \frac{{2n+6 \choose n+3} / (n+3)}{{2n \choose n}/(n+1)} \sim \frac{5}{4} 4^3 = 80.$$

% This is in agreement with the asymptotic result provided in the second row of the right-most column of Table~3 in \cite{Rosenberg13:tcbb}, where the size is
% defined as the total number of taxa. In fact, $t^{(n)}$ has $n+4$ taxa and the result of the mentioned table can be thus expressed as $$h_n \sim \frac{5}{4} % c_{(n+4)-1}.$$
% Observe that by eq.~(\ref{lucia3}) we have the asymptotic equivalence  $$\frac{5}{4} c_{n+3} \sim \frac{5}{4} \frac{4^{n+3}}{n^{3/2} \sqrt{\pi}} = 80
% \frac{4^n}{n^{3/2} \sqrt{\pi}} \sim 80 c_n,$$ where the latter is the asymptotic determined by our method.

%%%%%%%%%%%%%%%%%%%%%%%%%%%%%%%%%%%%%%%%%%%%%%%%%%%%%%%%%
%%%%%%%%%%%%%%%%%%%%%%%%%%%%%%%%%%%%%%%%%%%%%%%%%%%%%%%%%
%%%%%%%%%%%%%%%%%%%%%%%%%%%%%%%%%%%%%%%%%%%%%%%%%%%%%%%%%

\subsubsection{Determining $\beta_t$ from the seed tree $t$}
\label{secConnection}

We have shown in Corollary \ref{pluto} that the number of matching coalescent histories $h_n$ for the caterpillar-like family $t^{(n)}$ is, for a constant $\beta_t$, asymptotic to $\beta_t c_n$. We can now assemble our results to describe a procedure that given a seed tree $t$ with $|t| \geq 2$ determines both the generating function with coefficients $h_n$ and the constant $\beta_t$.

\begin{enumerate}
\item[(i)] Determine by eq.~(\ref{peze}) the polynomial $e_{t,k}$ in $k \geq 0$ that counts the number of $k$-extended histories of $t$.
\item[(ii)] Compute from eq.~(\ref{clinclin}) the polynomial in $m$ that counts for $m \geq 1$ the number of $m$-rooted histories of $t$.
\item[(iii)] Obtain the generating function $g(y) = \sum_{m=1}^\infty h_{0,m} y^m$ with coefficients $h_{0,m}$ by using Proposition~\ref{pinco}.
\item[(iv)] Determine the generating function $f(z)=\sum_{n=0}^\infty h_n z^n$ with coefficients $h_n$ by applying Proposition~\ref{pallino}.
\item[(v)] Write $g(y)$ as a linear combination according to eq.~(\ref{combin}), determining the values of $b$, $J$, and the $a_j$ and $q_j$.
\item[(vi)] Compute the asymptotic constant $\beta_t$ from eq.~(\ref{be2}).
\end{enumerate}

We have programmed this procedure in Mathematica; starting from a given seed tree $t$, our program \texttt{CatFamily.nb} can automatically compute for the caterpillar-like family $t^{(n)}$ the generating function with coefficients $h_n$ and the asymptotic constant $\beta_t$. Using this program, we have determined the value of $\beta_t$ for each seed tree with 9 taxa, collecting the results in Table~\ref{tavola1}.

Recall that Rosenberg \cite{Rosenberg13:tcbb} reported the asymptotic constant multiples of the Catalan numbers, $\beta_t^*$, which represent the asymptotic numbers of coalescent histories for seed trees with up to 8 taxa, indexing the results by the number of taxa $m$ rather than by the index $n$ of the caterpillar-like family. Also recall that for seed tree $t$, tree $t^{(n)}$ has $m=|t|+n$ taxa (Fig.~\ref{figFamilies2a1}). In the notation of \cite{Rosenberg13:tcbb}, writing $A_{t_{m},1}$ as the number of matching coalescent histories in the caterpillar-like tree with seed tree $t$ and $m \geq |t|$ taxa, we have $h_n = A_{t_m ,1}.$

By eq.~(\ref{lucia3}), we have the asymptotic equivalence $c_n \sim c_{n+k}/4^{k}$ for each positive integer $k$. Therefore,
\begin{equation}\label{stella}
A_{t_m,1} = h_n \sim \beta_t c_n \sim \frac{\beta_t}{4^{|t|-1}} c_{n+|t|-1} = \beta_t^{*} c_{m-1},
\end{equation}
where the asymptotic constant $\beta_t$ of Corollary~\ref{pluto} is normalized to obtain
\begin{equation}\label{stellina}
\beta_t^{*} = \frac{\beta_t}{4^{|t|-1}}.
\end{equation}
This computation converts the asymptotic constant multiple $\beta_t$ of $c_n$ into a corresponding multiple $\beta_t^*$ of $c_{m-1}$, as reported in \cite{Rosenberg13:tcbb} for small trees. Comparing Table \ref{tavola1} with Table 3 of \cite{Rosenberg13:tcbb}, we see that for the cases examined by \cite{Rosenberg13:tcbb}, the values of $\beta_t^*$ we compute from the associated $\beta_t$ agree with the values that were previously reported.

%%%%%%%%%%%%%%%%%%%%%%%%%%%%%%%%%%%%%%%%%%%%%%%%%%%%%%%%%
%%%%%%%%%%%%%%%%%%%%%%%%%%%%%%%%%%%%%%%%%%%%%%%%%%%%%%%%%
%%%%%%%%%%%%%%%%%%%%%%%%%%%%%%%%%%%%%%%%%%%%%%%%%%%%%%%%%
\begin{table}
\vspace{-.2cm}
\caption{{\small Asymptotic constants $\beta_t$ with $h_n \sim \beta_t c_n$, for seed trees $t$ with 9 taxa.}}
\label{tavola1}
\fontsize{9}{11}\selectfont
\begin{center}
\begin{tabular}{| c | c | c || c | c | c |}\hline \label{tavola}
Seed tree $t$ & $\beta_t$ & $\beta_t^{*}$ & Seed tree $t$ & $\beta_t$ & $\beta_t^{*}$ \\[2ex] \hline
& & & & & \\
\includegraphics[scale=.17,bb=200 200 210 250]{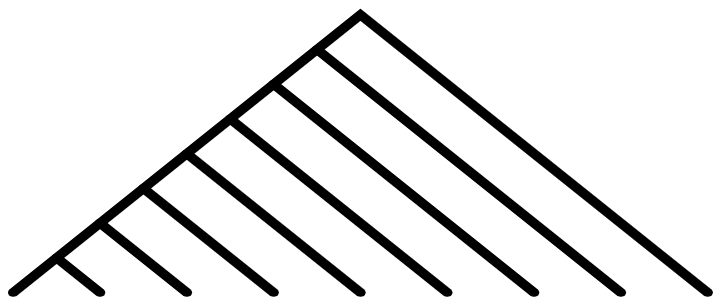}       & 65,536  & 1
& \includegraphics[scale=.17,bb=200 200 210 250]{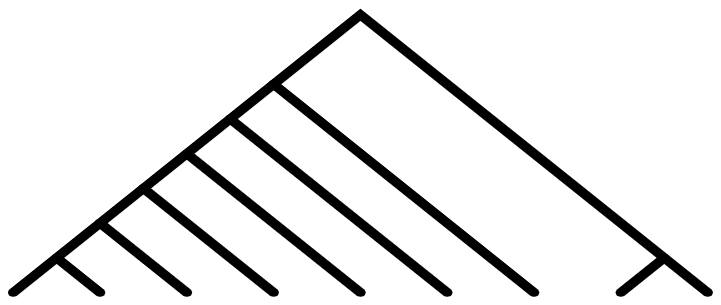}    & 128,864 & 4,027/2,048 \\[2ex]
\includegraphics[scale=.17,bb=200 200 210 250]{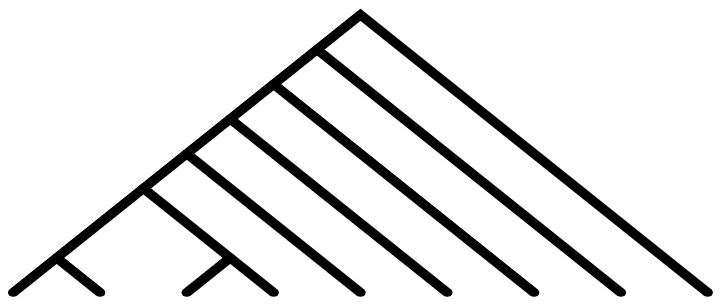}       & 81,920  & 5/4
& \includegraphics[scale=.17,bb=200 200 210 250]{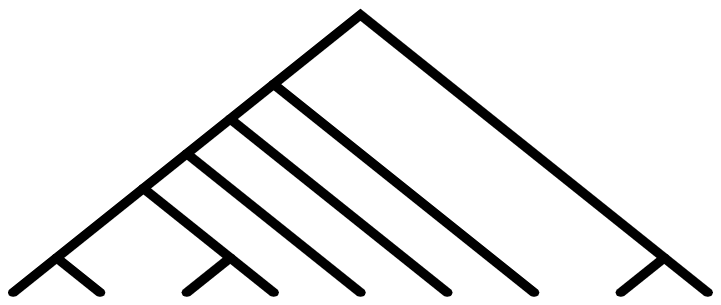}    & 166,624 & 5,207/2,048 \\[2ex]
\includegraphics[scale=.17,bb=200 200 210 250]{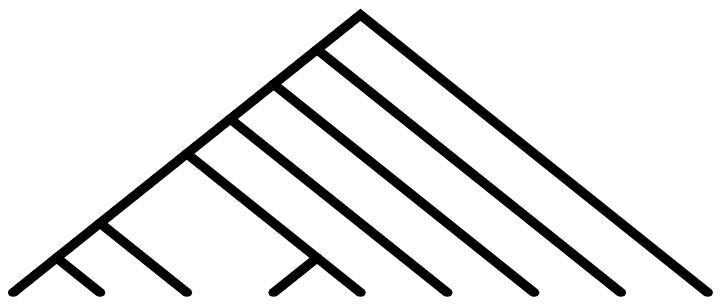}       & 94,208  & 23/16
& \includegraphics[scale=.17,bb=200 200 210 250]{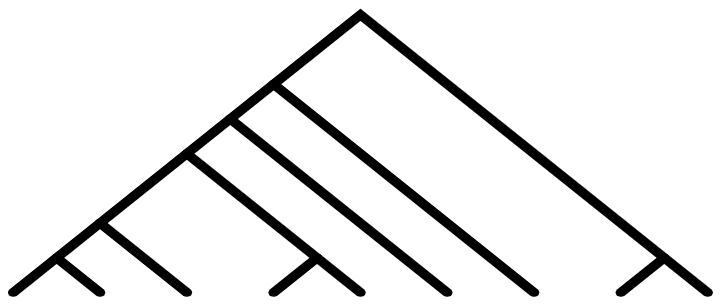}    & 197,296 & 12,331/4,096 \\[2ex]
\includegraphics[scale=.17,bb=200 200 210 250]{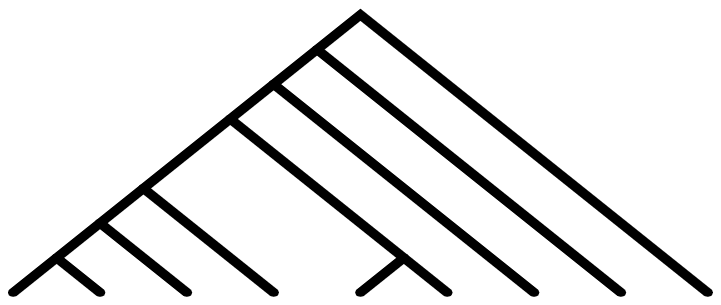}       & 104,448 & 51/32
&  \includegraphics[scale=.17,bb=200 200 210 250]{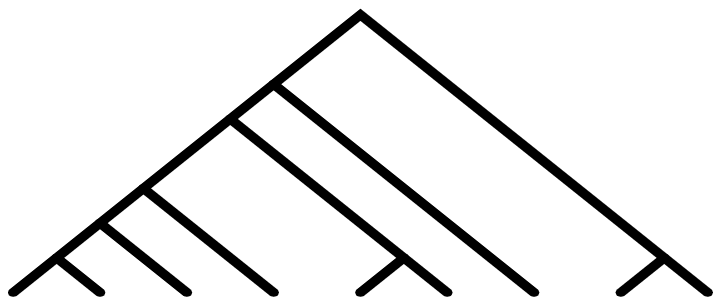}   & 224,704 & 3,511/1,024 \\[2ex]
\includegraphics[scale=.17,bb=200 200 210 250]{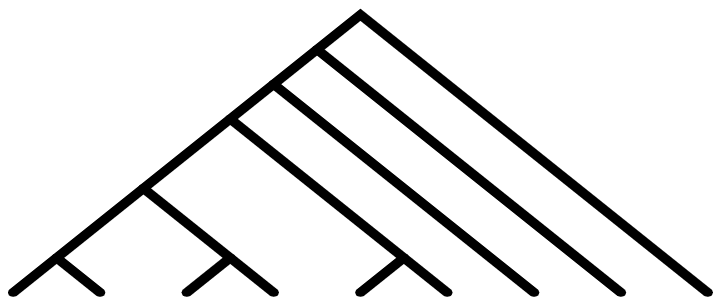}       & 138,240 & 135/64
&  \includegraphics[scale=.17,bb=200 200 210 250]{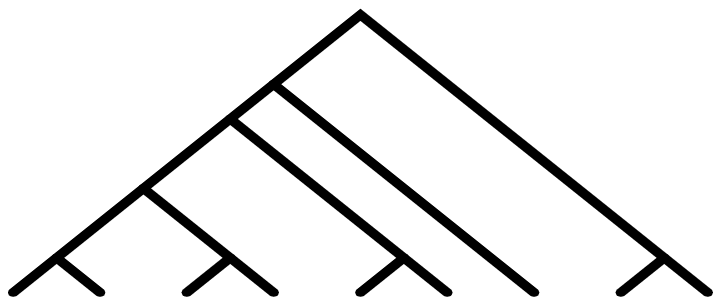}   & 308,576 & 9,643/2,048 \\[2ex]
\includegraphics[scale=.17,bb=200 200 210 250]{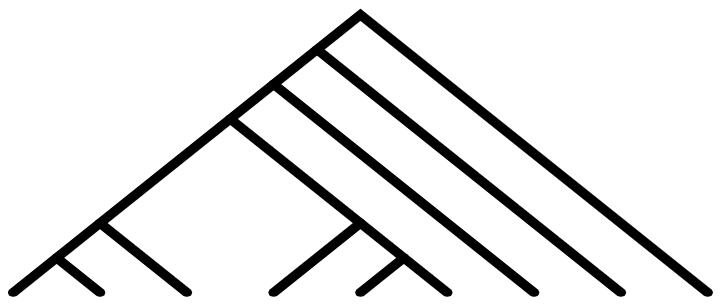}       & 118,784 & 29/16
&   \includegraphics[scale=.17,bb=200 200 210 250]{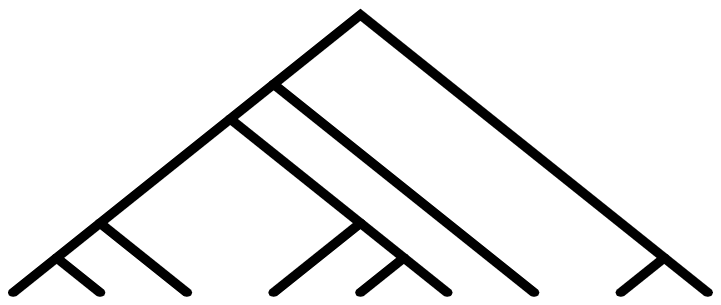}  & 262,000 & 16,375/4,096 \\[2ex]
\includegraphics[scale=.17,bb=200 200 210 250]{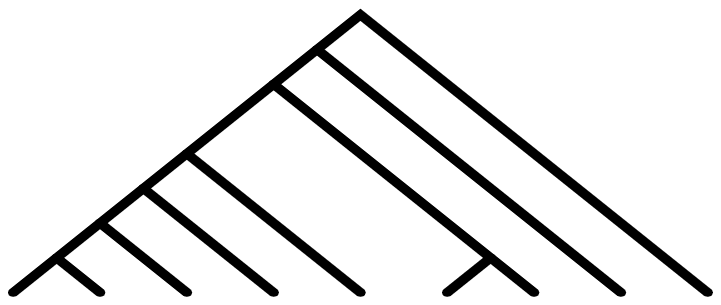}       & 113,408 & 443/256
&  \includegraphics[scale=.17,bb=200 200 210 250]{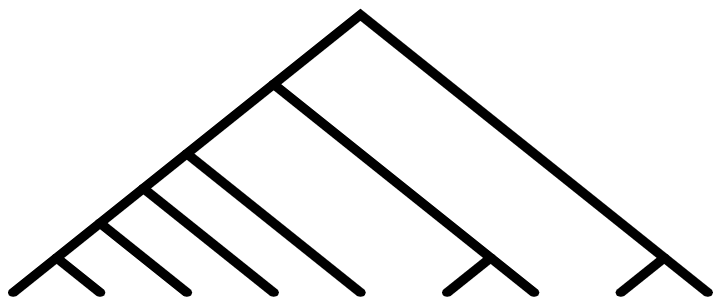}   & 250,272 & 7,821/2,048 \\[2ex]
\includegraphics[scale=.17,bb=200 200 210 250]{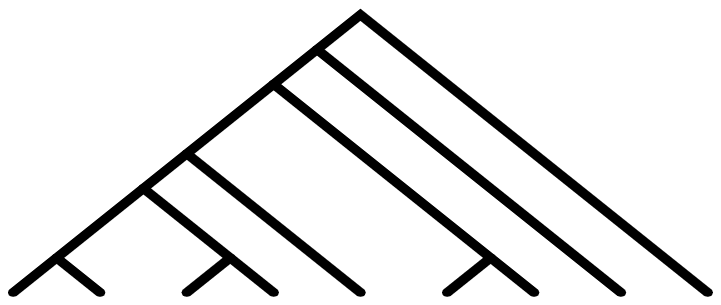}       & 148,480 & 145/64
&  \includegraphics[scale=.17,bb=200 200 210 250]{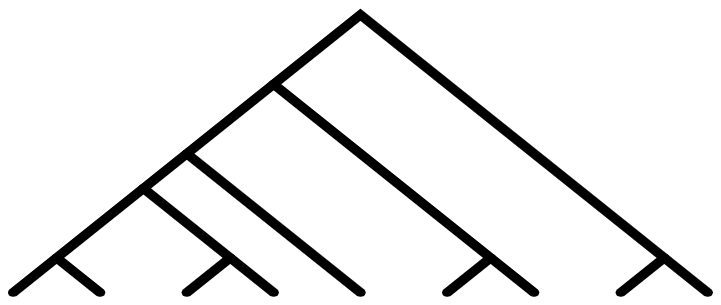}   & 339,504 & 21,219/4,096 \\[2ex]
\includegraphics[scale=.17,bb=200 200 210 250]{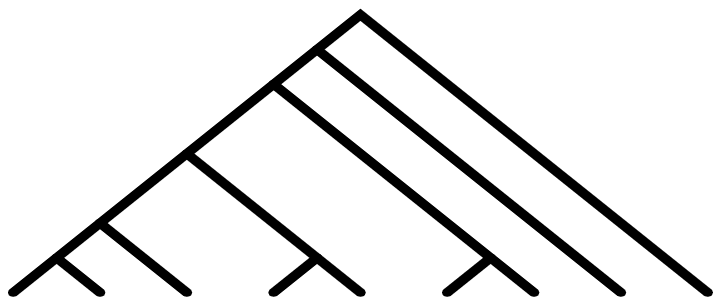}       & 177,664 & 347/128
&  \includegraphics[scale=.17,bb=200 200 210 250]{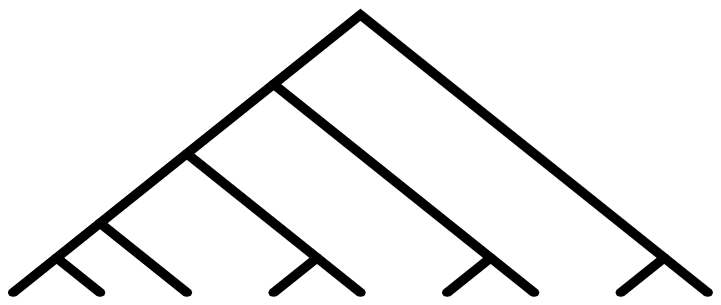}   & 417,632 & 13,051/2,048 \\[2ex]
\includegraphics[scale=.17,bb=200 200 210 250]{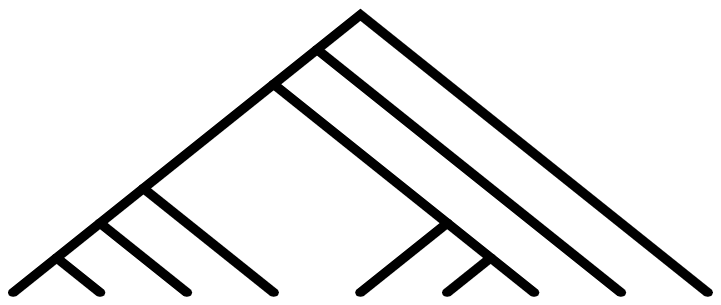}      & 141,312 & 69/32
&  \includegraphics[scale=.17,bb=200 200 210 250]{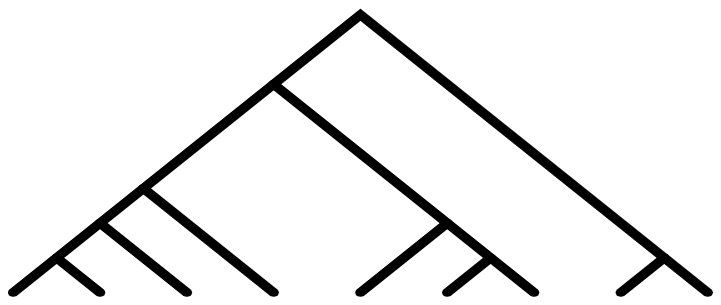}   & 326,240 & 10,195/2,048 \\[2ex]
\includegraphics[scale=.17,bb=200 200 210 250]{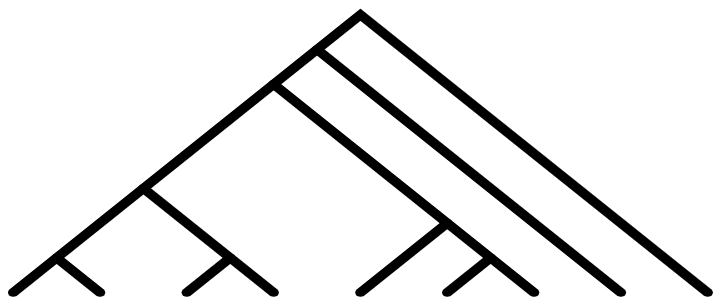}      & 193,536 & 189/64
&  \includegraphics[scale=.17,bb=200 200 210 250]{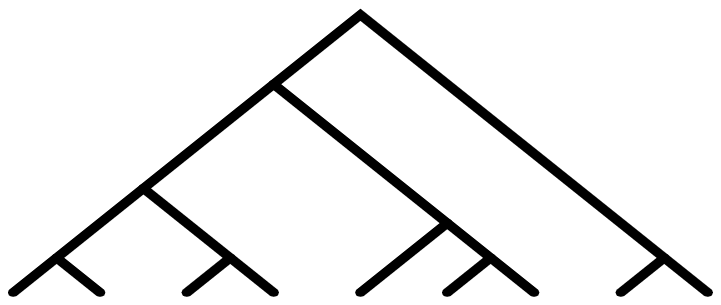}   & 464,128 & 1,813/256 \\[2ex]
\includegraphics[scale=.17,bb=200 200 210 250]{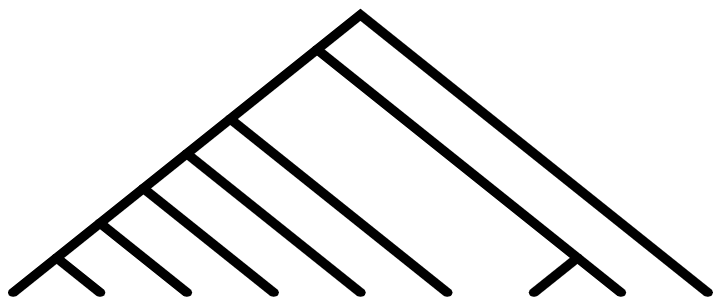}      & 121,472 & 949/512
&  \includegraphics[scale=.17,bb=200 200 210 250]{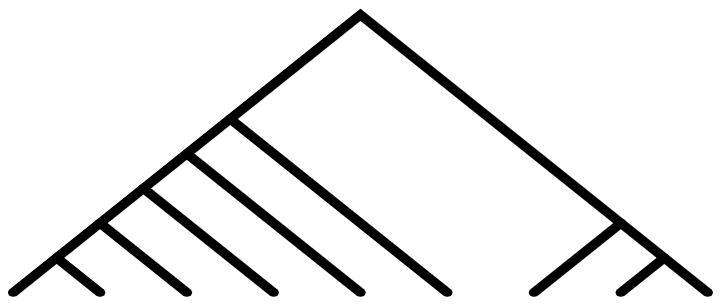}   & 182,912 & 1,429/512 \\[2ex]
\includegraphics[scale=.17,bb=200 200 210 250]{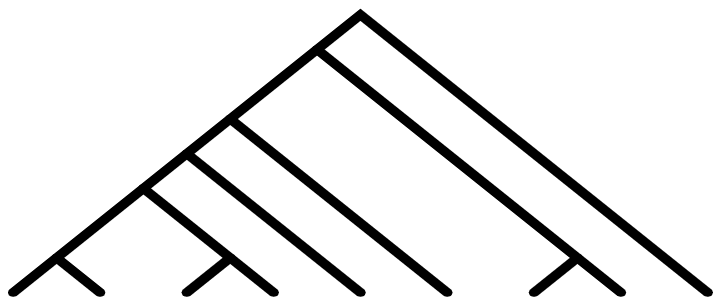}      & 157,888 & 2,467/1,024
&   \includegraphics[scale=.17,bb=200 200 210 250]{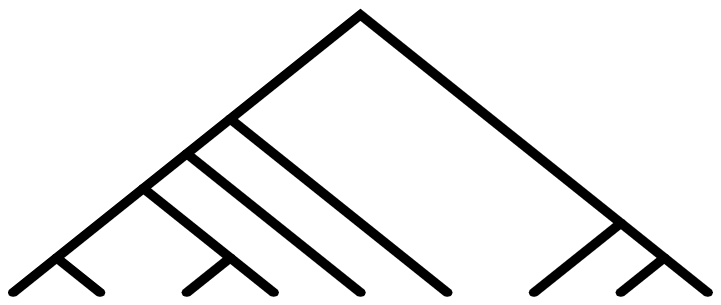}  & 243,904 & 3,811/1,024 \\[2ex]
\includegraphics[scale=.17,bb=200 200 210 250]{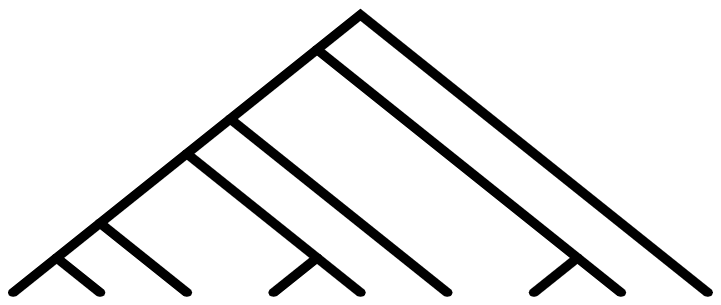}      & 187,776 & 1,467/512
&  \includegraphics[scale=.17,bb=200 200 210 250]{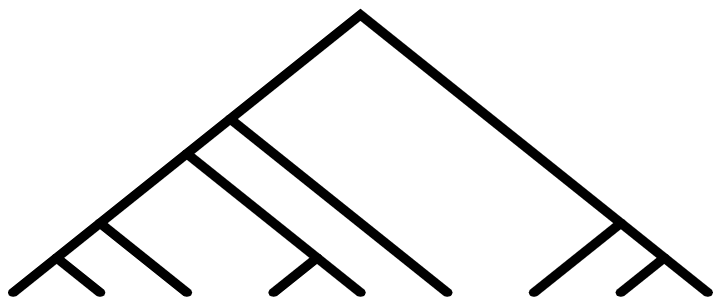}   & 296,064 & 2,313/512 \\[2ex]
\includegraphics[scale=.17,bb=200 200 210 250]{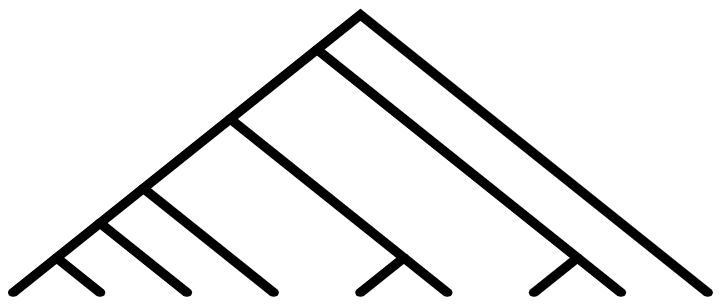}      & 214,720 & 3,355/1,024
&  \includegraphics[scale=.17,bb=200 200 210 250]{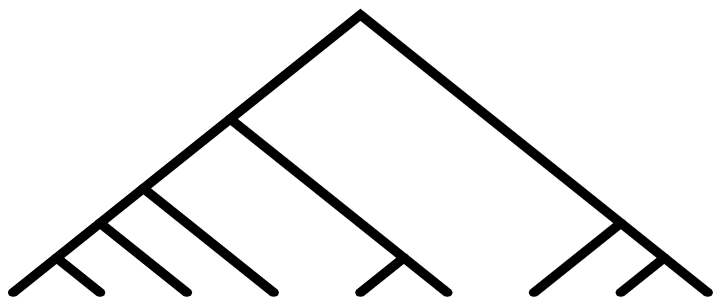}   & 344,512 & 5,383/1,024 \\[2ex]
\includegraphics[scale=.17,bb=200 200 210 250]{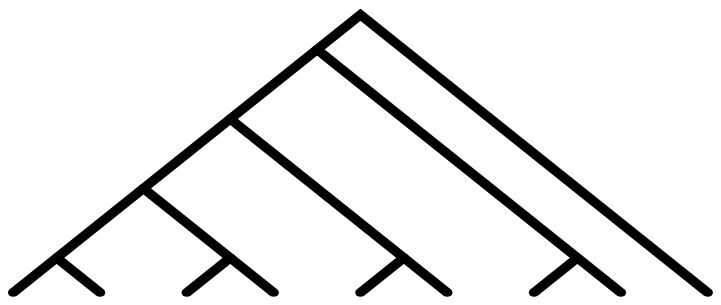}      & 296,192 & 1,157/256
&   \includegraphics[scale=.17,bb=200 200 210 250]{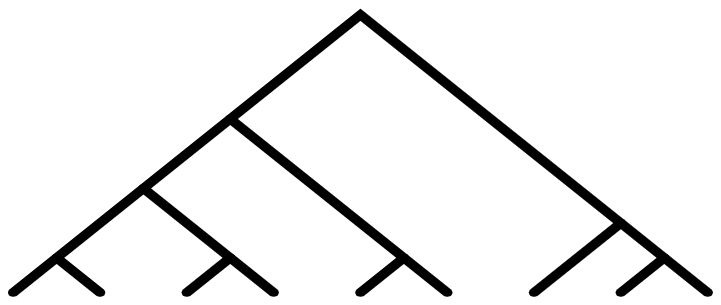}  & 487,808 & 3,811/512 \\[2ex]
\includegraphics[scale=.17,bb=200 200 210 250]{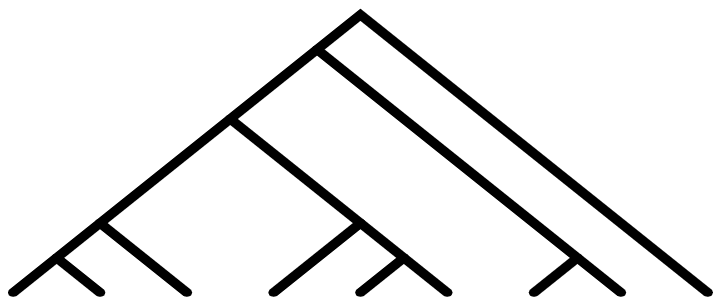}      & 251,136 & 981/256
&  \includegraphics[scale=.17,bb=200 200 210 250]{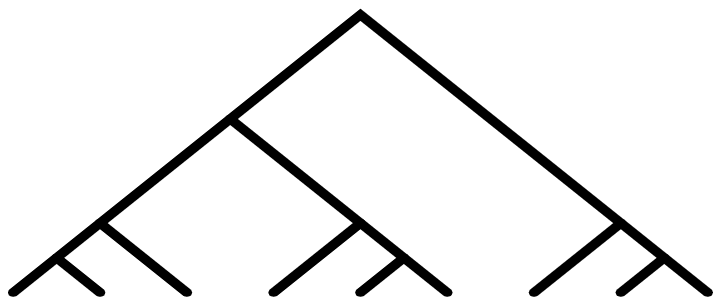}   & 410,112 & 801/128 \\[2ex]
\includegraphics[scale=.17,bb=200 200 210 250]{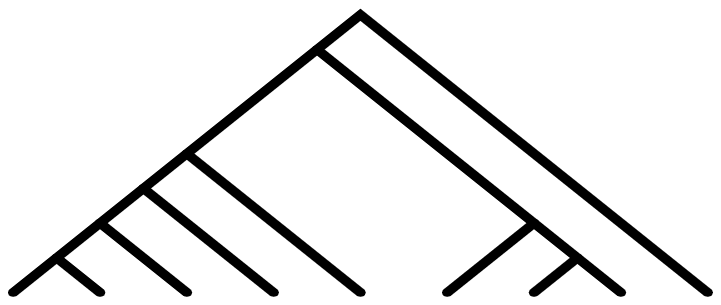}      & 162,560 & 635/256
&    \includegraphics[scale=.17,bb=200 200 210 250]{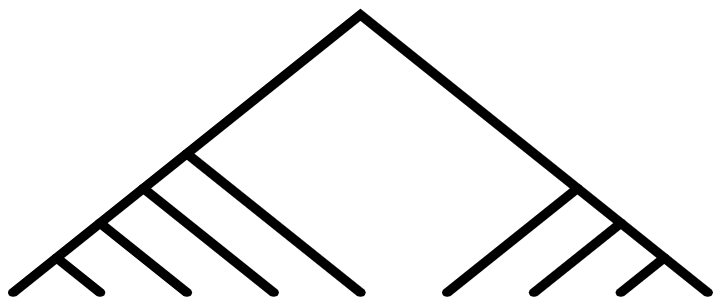} & 214,016 & 209/64 \\[2ex]
\includegraphics[scale=.17,bb=200 200 210 250]{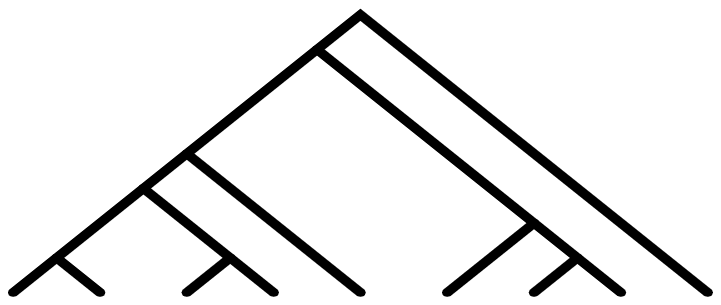}      & 219,136 & 107/32
&  \includegraphics[scale=.17,bb=200 200 210 250]{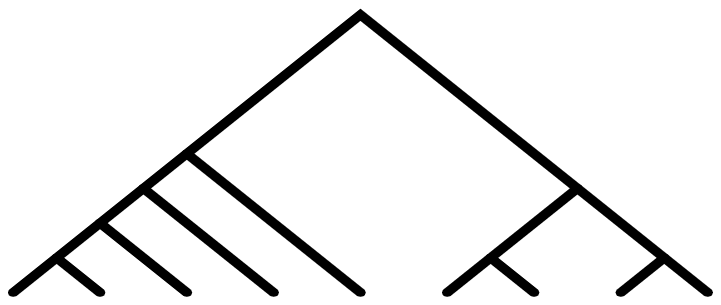}   & 306,112 & 4,783/1,024 \\[2ex]
\includegraphics[scale=.17,bb=200 200 210 250]{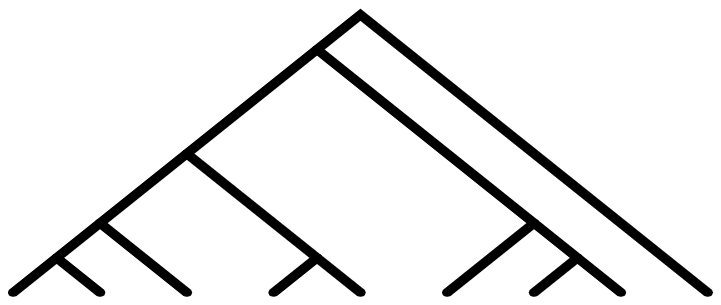}      & 268,288 & 131/32
&  \includegraphics[scale=.17,bb=200 200 210 250]{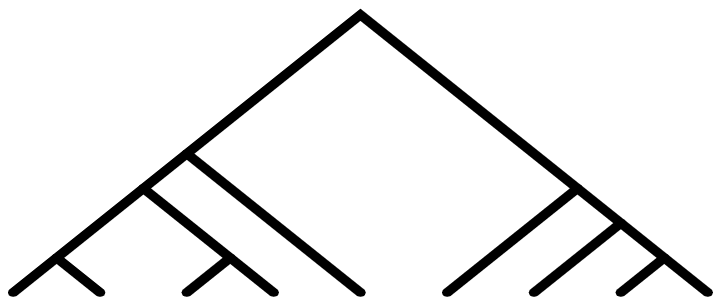}   & 294,784 & 2,303/512 \\[2ex]
\includegraphics[scale=.17,bb=200 200 210 250]{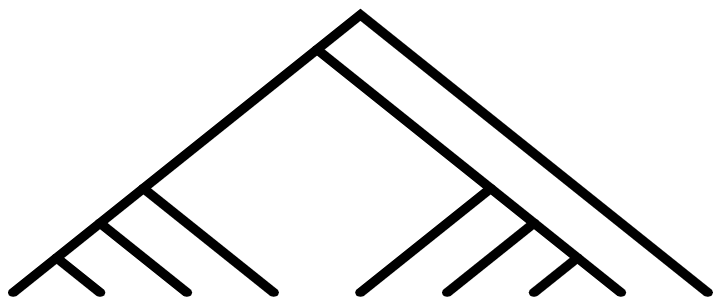}      & 177,664 & 347/128
&  \includegraphics[scale=.17,bb=200 200 210 250]{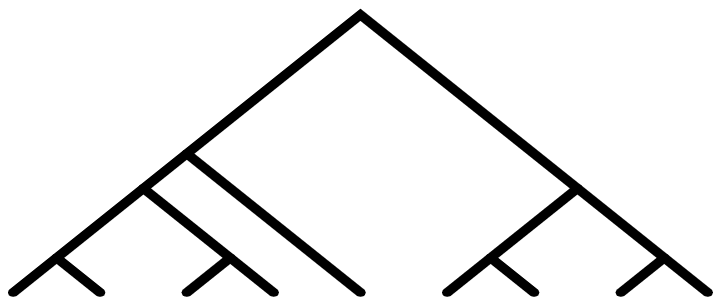}   & 425,216 & 1,661/256 \\[2ex]
\includegraphics[scale=.17,bb=200 200 210 250]{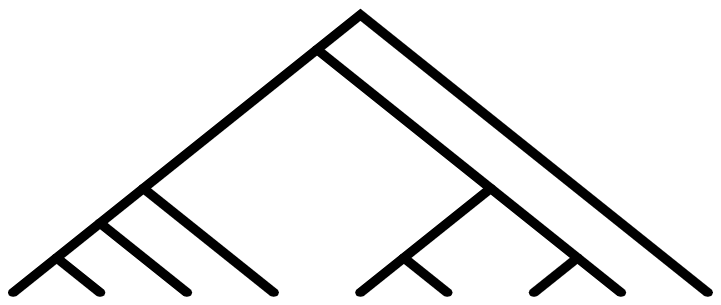}      & 249,344 & 487/128
&   \includegraphics[scale=.17,bb=200 200 210 250]{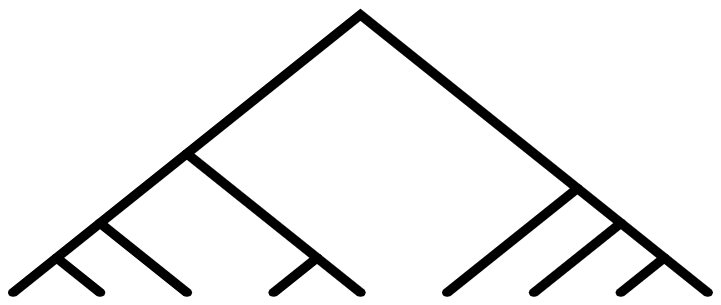}  & 366,720 & 2,865/512 \\[2ex]
\includegraphics[scale=.17,bb=200 200 210 250]{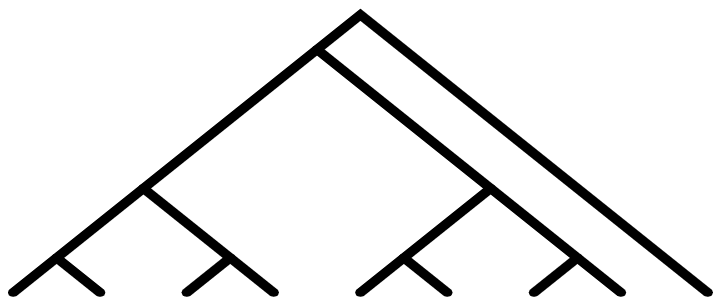}      & 353,536 & 1,381/256
&  \includegraphics[scale=.17,bb=200 200 210 250]{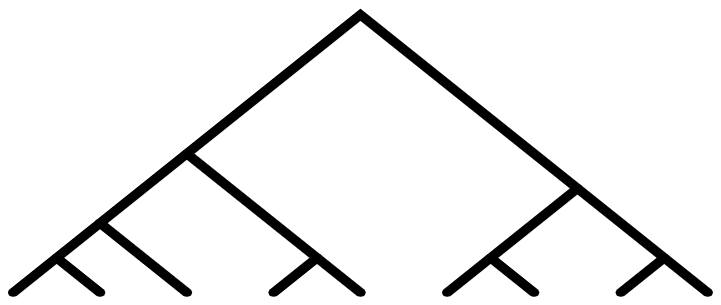}   & 532,224 & 2,079/256 \\[2ex]
\hline
\end{tabular}
\end{center}
\vspace{-.1cm}
{\small{Values of $\beta_t$ appear for each of the 46 unlabeled species trees with 9 taxa. For each species tree $t$, we also provide the constant $\beta_t^{*}=\beta_t/4^8$ (eq.~(\ref{stellina})). Trees are listed in increasing order by rank as defined in Section 2 of \cite{Rosenberg13:tcbb}. In the left column, each seed tree $t$ belongs to a caterpillar-like family $(\tilde{t}^{(n)})_n$, with $|\tilde{t}| < 9$. In these cases, we recover the values of $\beta_t^{*}$ as determined in Table 3 of \cite{Rosenberg13:tcbb}.}}
\end{table}
%%%%%%%%%%%%%%%%%%%%%%%%%%%%%%%%%%%%%%%%%%%%%%%%%%%%%%%%%
%%%%%%%%%%%%%%%%%%%%%%%%%%%%%%%%%%%%%%%%%%%%%%%%%%%%%%%%%
%%%%%%%%%%%%%%%%%%%%%%%%%%%%%%%%%%%%%%%%%%%%%%%%%%%%%%%%%

%%%%%%%%%%%%%%%%%%%%%%%%%%%%%%%%%%%%%%%%%%%%%%%%%%%%%%%%%
%%%%%%%%%%%%%%%%%%%%%%%%%%%%%%%%%%%%%%%%%%%%%%%%%%%%%%%%%
%%%%%%%%%%%%%%%%%%%%%%%%%%%%%%%%%%%%%%%%%%%%%%%%%%%%%%%%%

\section{Conclusions}
\label{secConclusions}

In this paper, we have solved a problem left open by \cite{Rosenberg13:tcbb} on determining the number of coalescent histories for gene trees and species trees that have a matching labeled topology and that belong to a generic caterpillar-like family. We have proven that for any seed tree $t$, the integer sequence $(h_n)_{n\geq 0}$, whose $n$th element represents the number of matching coalescent histories of the caterpillar-like tree $t^{(n)}$, grows asymptotically as a constant multiple of the Catalan numbers, that is, $h_n \sim \beta_t c_n$, where the constant term $\beta_t > 0$ depends on the shape of the seed tree $t$. Rosenberg \cite{Rosenberg13:tcbb} had previously obtained this result for seed trees with at most 8 taxa; here, by using a succession rule for recursive enumeration and then applying techniques of analytic combinatorics, we have not only proven the existence of the constant $\beta_t$ for seed trees of any size, we have also produced a procedure that computes the constant $\beta_t$, as well as the expression for the generating function of the integers $(h_n)_{n\geq 0}$.

The numerical results on the constants $\beta_t$ extend the empirical observation of \cite{Rosenberg13:tcbb} that the caterpillar-like families that produce the largest numbers of matching coalescent histories are those whose seed tree has a high level of balance. By extending from seed trees with $|t| \leq 8$ taxa to those with $|t|=9$, we observe that the constants $\beta_t$ for the caterpillar-like families with the largest and smallest numbers of matching coalescent histories become further separated, so that for $n$ large, many more coalescent histories exist by which a gene tree can match the species tree for some species trees than for others. For the 9-taxon seed tree with the largest $\beta_t^*$, $\beta_t^* \approx 8.12$ compared to $\beta_t=1$ for the seed tree with the smallest $\beta_t^*$. Our procedure for evaluating $\beta_t$ and $\beta_t^*$ as a function of the seed tree can now enable further systematic analyses of the correlates of the constants $\beta_t$ and $\beta_t^*$, to facilitate additional explorations of determinants of the numbers of matching coalescent histories.

Nevertheless, although the constants $\beta_t$ and $\beta_t^*$ do depend on the seed tree, we have shown that all caterpillar-like families are asymptotically equivalent in their numbers of matching coalescent histories up to a constant factor. Thus, in considering large trees, the many caterpillar branches contribute to the asymptotic growth behavior of the number of matching coalescent histories---which follows a multiple of the Catalan numbers---and the seed tree contributes only to the constant by which the Catalan numbers are multiplied. From the viewpoint of computational complexity in evaluating gene tree probabilities according to formulas that sum over matching coalescent histories \cite{DegnanAndSalter05}, all caterpillar-like families have the same growth pattern up to a constant.

The extent to which other tree families follow the Catalan sequence in their numbers of matching coalescent histories remains unknown, though we have recently found a family, the \emph{lodgepole} family, for which the number of matching coalescent histories grows faster than with a constant multiple of the Catalan numbers \cite{DisantoAndRosenberg15}. The use of our substantially different approach employing analytic combinatorics opens new methods for theoretical analysis of coalescent histories and can potentially assist in understanding when Catalan-like growth, the rapid growth of the lodgepole family, and intermediate or perhaps still faster growth patterns will apply.

%%%%%%%%%%%%%%%%%%%%%%%%%%%%%%%%%%%%%%%%%%%%%%%%%%%%%%%%%
%%%%%%%%%%%%%%%%%%%%%%%%%%%%%%%%%%%%%%%%%%%%%%%%%%%%%%%%%
%%%%%%%%%%%%%%%%%%%%%%%%%%%%%%%%%%%%%%%%%%%%%%%%%%%%%%%%%

\section*{Appendix 1. The equation satisfied by $F(y,z)$}

In this appendix, we complete the derivation of eq.~(\ref{scarpa}) satisfied by $F(y,z)$. In the generating function $F(y,z)$ (eq.~(\ref{f})), each monomial $z^n y^m$ corresponds to a label $(n,m) \in L_n$ that in turn represents an $m$-rooted history of $t^{(n)}$. Recall that the multisets of labels $L_0, L_1, L_2, \dots$ (eq.~(\ref{babbino})) can be iteratively generated according to eq.~(\ref{dopo}) through the operator $\Omega$ defined in eq.~(\ref{omeg2}), starting from the multiset $L_0$. Also recall that by considering the multiset of labels $L = \cup_{n=0}^\infty L_n$, we can write $F(y,z) = \sum_{(n,m) \in L} z^n y^m.$ We use the iterative generation of the family of multisets $(L_n)_{n \geq 0}$ to obtain an equation for $F$.
% Following a standard approach \cite{BarcucciEtAl99}.

By eq.~(\ref{omeg2}), for $n \geq 0$ and $m \geq 2$, for each occurrence in $L_n$ of a label $(n,m)$, a copy of each label in set
$$\Omega\big( (n,m) \big) = \{(n+1,m+j) : j \geq -1 \}$$
belongs to the multiset $L_{n+1}$. Thus, in algebraic terms, each time that an expression $z^n y^m$ with $n \geq 0$ and $m \geq 2$ is counted in the generating function $F$---written $z^n y^m \in F$ in what follows---the terms
%\begin{equation}
%\label{termini1}
$z^{n+1} \sum_{j=m-1}^\infty y^j$
%\end{equation}
appear in $F$ as well.
% In symbols we write
% \begin{equation}
% z^ny^m \in F (n \geq 0, m\geq 2 ) \rightarrow z^{n+1}(y^{m-1}+y^{m}+y^{m+1}+ \ldots) \in F.
% \end{equation}
Summing over all possible $z^n y^m \in F$ with $n \geq 0$ and $m \geq 2$, we obtain
\begin{eqnarray}
\sum_{z^n y^m \in F \,:\, n\geq0, m \geq 2} \bigg( z^{n+1} \sum_{j=m-1}^\infty y^j \bigg)
& = & \frac{z}{y} \sum_{z^n y^m \in F \,:\, n\geq0, m \geq 2} \bigg( z^n y^m \sum_{j=0}^\infty y^j \bigg). \label{pino}
% \sum_{z^n y^m \in F \,:\, n\geq0, m \geq 2} z^n(y^m+y^{m+1}+y^{m+2}+\ldots)
% \sum_{z^n y^m \in F \,:\, n\geq0, m \geq 2} z^{n+1}(y^{m-1}+y^{m}+y^{m+1}+\ldots) &=&
% z/y \cdot \sum_{z^n y^m \in F \,:\, n\geq0, m \geq 2} z^n(y^m+y^{m+1}+y^{m+2}+\ldots) \\\label{pino}
% &=& z/y \cdot \sum_{z^n y^m \in F \,:\, n\geq0, m \geq 2} z^ny^m(1+y^{1}+y^{2}+\ldots).
\end{eqnarray}
% whose terms are those generated by those $z^ny^m$ with $n \geq 0$ and $m\geq 2$ counted in $F$.
% a which represents those labels $\ell \in L$ such that there exists a label $(n,m) \in L$ with $n \geq 0$, $m \geq 2$ and  $\ell \in \Omega\big( (n,m)
% \big)$.

Similarly, for $n \geq 0$ and $m = 1$, for each occurrence in $L_n$ of a label $(n,1)$, a copy of each label in set $\Omega\big( (n,1) \big) = \{(n+1,j) :j \geq 1 \}$ appears in the multiset $L_{n+1}$. Thus, for each term $z^n y \in F$, with $n \geq 0$, the terms $z^{n+1} \sum_{j=1}^\infty y^j$ are counted in $F$ as well. Summing these terms for all $z^n y \in F$ with $n \geq 0$, we obtain
\begin{equation}
\label{rino}
\sum_{z^n y  \in F \,:\, n \geq 0} \bigg( z^{n+1} \sum_{j=1}^\infty y^j \bigg) = zy  \sum_{z^n y \in F \,:\, n \geq 0} \bigg(z^n \sum_{j=0}^\infty y^j \bigg).
\end{equation}
% whose terms are those generated by those $z^ny^m$ with $n \geq 0$ and $m = 1$ counted in $F$.
% which is the algebraic expression that represents those labels $\ell$ such that there exists a label $(n,m)$ with $n \geq 0$, $m =1 $ and  $\ell \in
% \Omega\big( (n,m)  \big)$.

Notice that the sum of the expressions in eqs.~(\ref{pino}) and (\ref{rino}) is the algebraic representation of the multiset of labels $L \setminus L_0$. More precisely, each term $z^n y^m \in F$ associated with a label $(n,m) \in L_n$, with $n \geq 1$, is counted---and counted exactly once---in the sum of eqs.~(\ref{pino}) and (\ref{rino}).
% Rooted histories are constructed via the operator $\Omega$. The iterative application of the rules $\Omega(n,m)$ (\ref{omeg}) can be translated, as detailed % in \cite{bosc}, in terms of two sums, one for each case $m \geq 2$ or $m=1$.
% More precisely, the correspondance is given as follows:
% \begin{eqnarray}\nonumber
% \text{$\forall n\geq 0$ and $\forall m \geq 2$, }\Omega(n,m) =  \{(n+1,m+j)\}_{j \geq -1} & \Rightarrow & \sum_{z^n y^m \,:\, n\geq0, m \geq 2}
% z^{n+1}(y^{m-1}+y^{m}+y^{m+1}...) \\\label{pino}
% & = & z/y \cdot \sum_{z^n y^m \,:\, n\geq0, m \geq 2} z^n(y^m+y^{m+1}+...) \\\nonumber
% \text{$\forall n\geq 0$ and $m = 1$, }\Omega(n,m) =  \{(n+1,j)\}_{j \geq 1} & \Rightarrow & \sum_{z^n y^m \,:\, n \geq 0,m = 1} z^{n+1}(y^{1}+y^{2}+y^{3}...) % \\\label{rino}
% & = & zy \cdot \sum_{z^n y^m \,:\, n \geq 0,m = 1} z^n(1+y^{1}+...) .
% \end{eqnarray}
% Note how each sum is considered over the terms $z^ny^m$, with $n$ and $m$ given as for the rooted histories $(n,m)$ to which we apply the operator $\Omega$
% in each case $m\geq 2$ and $m=1$. Moreover, observe the correspondance between the exponents in the argument of each sum and the set of labels produced as
% $\Omega(n,m)$.
Therefore, to complete the description of $F$, we require only those terms $z^0 y^m$ associated with labels $(0,m) \in L_0$. These terms are represented by
\begin{equation}\label{gino}
\sum_{(0,m) \in L_0} z^0 y^m = \sum_{m=1}^\infty h_{0,m} y^m = g(y),
\end{equation}
considering that $h_{0,m} = |\{ \ell \in L_0 : \ell = (0,m) \}$ (eq.~(\ref{ricordi})) and that by definition, $g(y)=\sum_{m=1}^\infty h_{0,m} y^m$ (eq.~(\ref{g})).

% The iterative application of the operator $\Omega$ needs a basic step made of the rooted histories of $t^{(0)}$, that is, those labelled with $(0,m)$, $m
% \geq 1$. Their algebraic counterpart is taken into account by the generating function $g$ defined as in (\ref{g}). The correspondance is then the following
% \begin{eqnarray}\label{gino}
% \{(0,m)\}_{m \geq 1} & \Rightarrow & \sum_{n=0,m\geq 1} z^n y^m = z^0 \sum_{m\geq 1}y^m = g(y).
% \end{eqnarray}

% Because each label $(n,m) \in L$ belongs either to $L_0$  -- and then it is considered in $g(y)$ (\ref{gino}) -- or to $L_n$ for some $n \geq 1$ -- and then % it is counted (only once) in the sum of (\ref{pino}) and (\ref{rino}) --
We can now equate the full generating function $F(y,z)$ to the sum of eqs.~(\ref{gino}), (\ref{pino}), and (\ref{rino}), obtaining
\begin{equation}
%\label{scarpetta}
F(y,z) = g(y) + \frac{z}{y} \sum_{z^n y^m \in F \,:\, n\geq0, m \geq 2} \bigg( z^n y^m \sum_{j=0}^\infty y^j \bigg)
+ zy \sum_{z^n y \in F \,:\, n \geq 0} \bigg(z^n \sum_{j=0}^\infty y^j \bigg).
\end{equation}
Applying the fact that $\sum_{j=0}^\infty y^j = 1/(1-y)$ for $y$ near 0 in the complex plane, we then have
\begin{equation}\label{scarpuccia}
F(y,z) = g(y) + \frac{z}{y(1-y)} \bigg( \sum_{z^n y^m \in F \,:\, n \geq 0, m \geq 2} z^n y^m \bigg) + \frac{z y}{1-y}\bigg( \sum_{z^n y \in F \,:\, n \geq 0} z^n \bigg).
\end{equation}

By eq.~(\ref{f}) and the fact that the multisets $L_n$ of labels $(n,m)$ for $m$-rooted histories of $t^{(n)}$ have $h_{n,m}$ elements,
\begin{eqnarray}
\sum_{z^n y \in F \,:\, n\geq 0} z^n                  & = & \frac{\partial F}{\partial y}(0,z) \nonumber \\
\sum_{z^n y^m \in F \,:\, n \geq 0, m \geq 2} z^n y^m & = & \bigg(\sum_{z^n y^m \in F  \,:\, n \geq 0, m \geq 1} z^n y^m \bigg) - \bigg(\sum_{z^n y \in F \,:\, n \geq 0} z^n y \bigg) = F(y,z) - y\frac{\partial F}{\partial y}(0,z). \nonumber
\end{eqnarray}
Substituting in eq.~(\ref{scarpuccia}), the last two expressions yield
\begin{equation}
\label{scarpina}
F(y,z) = g(y) + \frac{z}{y(1-y)} \bigg( F(y,z) - y \frac{\partial F}{\partial y}(0,z) \bigg) +  \frac{z y}{1-y} \frac{\partial F}{\partial y}(0,z),
\end{equation}
which can be rewritten as in eq.~(\ref{scarpa}).

%%%%%%%%%%%%%%%%%%%%%%%%%%%%%%%%%%%%%%%%%%%%%%%%%%%%%%%%%
%%%%%%%%%%%%%%%%%%%%%%%%%%%%%%%%%%%%%%%%%%%%%%%%%%%%%%%%%
%%%%%%%%%%%%%%%%%%%%%%%%%%%%%%%%%%%%%%%%%%%%%%%%%%%%%%%%%

\section*{Appendix 2. The dominant singularity and singular expansion of $\tilde{f}(z)$}

This appendix obtains the singular expansion of $\tilde{f}(z)$ described in eq.~(\ref{asino}). In eq.~(\ref{titti}), we have defined $\tilde{f}(z)$ as a composition $\tilde{f}(z) = g(Y(z))$, with the internal function $Y(z)$ as in eq.~(\ref{ker}) and the external function $g(y)$ as in eq.~(\ref{combin}). Owing to the presence of the square root in the expression for $Y(z)$, the dominant singularity of the internal function $Y(z)$---the singularity nearest the origin of the complex plane---is at $z=\frac{1}{4}$. Computing the value of $Y(z)$ at its dominant singularity, we obtain $Y(\frac{1}{4})= \frac{1}{2}$. In particular, we have $Y(\frac{1}{4}) < 1$, where 1 is the radius of convergence of the finite series corresponding to the external function $g$ in $\tilde{f}$. Indeed, it immediately follows from Proposition~\ref{pinco} that $y=1$ is the dominant singularity of $g(y)$.

As detailed in Section VI.9 of \cite{FlajoletAndSedgewick09}, on dominant singularities of compositions, we are in the setting of the subcritical case, in which the inequality $Y(\frac{1}{4}) < 1$ implies that the dominant singularity of $g(Y(z))$ coincides with the dominant singularity $z=\frac{1}{4}$ of the internal function $Y(z)$ rather than the dominant singularity $y=1$ of the external function $g(y)$. The desired singular expansion of $\tilde{f}(z) = g(Y(z))$ at the dominant singularity $z=\frac{1}{4}$ can be obtained by inserting $y=Y(z)$ in the regular (non-singular) expansion of $g(y)$ at $y = Y(\frac{1}{4})=\frac{1}{2}$.

To recover the expansion of $g(y)$ at $y = \frac{1}{2}$, we expand and then sum each term $q_j [y^{a_j}/(1-y)^{b}]$ of the finite linear combination in eq.~(\ref{combin}). At $y = \frac{1}{2}$, each of these terms is an analytic function, and we can thus use Taylor's formula to produce the desired expansion. We obtain at $y = \frac{1}{2}$
$$q_j \frac{y^{a_j}}{(1-y)^{b}}  = 2^{b-a_j} q_j + 2^{b+1-a_j}(a_j+b)q_j\bigg( y-\frac{1}{2} \bigg)  \pm \mathcal{O}\bigg( \big( y-\frac{1}{2} \big)^2 \bigg). $$
By summing over the indices $1 \leq j \leq J$ of eq.~(\ref{combin}), the expansion of $g(y)$ at $y = \frac{1}{2}$ is
\begin{equation}
\label{numero}
g(y) = \alpha_t + \beta_t \bigg(y-\frac{1}{2} \bigg) \pm \mathcal{O}\bigg( \big( y-\frac{1}{2} \big)^2 \bigg),
\end{equation}
with the constants $\alpha_t$ and $\beta_t$ defined as in eqs.~(\ref{be1}) and (\ref{be2}).
% \begin{equation}\label{be}
% \alpha = \sum_{i=1}^J 2^{b-a_i} k_i \mathrm{\,\,\, and \,\,\, }
% \beta = \sum_{i=1}^J 2^{1+b-a_i}(a_i+b)k_i.
% \end{equation}
Plugging $y=Y(z)$ from eq.~(\ref{ker}) into eq.~(\ref{numero}), we finally obtain the singular expansion of $\tilde{f}(z)$ at $z=\frac{1}{4}$ as in eq.~(\ref{asino}).

% \begin{eqnarray}\nonumber
% \tilde{f}(z) &=& \alpha + \beta \cdot \left(-\sqrt{\frac{1}{4}-z}\right) \pm \mathcal{O}\left( \left(-\sqrt{\frac{1}{4}-z}\right)^2 \right)  \\\label{asino}
% &=& \alpha + \beta \cdot \left(-\frac{\sqrt{1-4z}}{2}\right) \pm \mathcal{O}(1-4z)  \\\nonumber
% & \sim &  \alpha + \beta \cdot \left(-\frac{\sqrt{1-4z}}{2}\right)
% \end{eqnarray}

%%%%%%%%%%%%%%%%%%%%%%%%%%%%%%%%%%%%%%%%%%%%%%%%%%%%%%%%%
%%%%%%%%%%%%%%%%%%%%%%%%%%%%%%%%%%%%%%%%%%%%%%%%%%%%%%%%%
%%%%%%%%%%%%%%%%%%%%%%%%%%%%%%%%%%%%%%%%%%%%%%%%%%%%%%%%%

\section*{Acknowledgments}
We acknowledge grant support from the National Science Foundation (DBI-1146722). A Mathematica notebook {\tt CatFamily.nb} implementing the procedure in Section \ref{secConnection} for obtaining from a seed tree $t$ the generating function $f(z)$, the coefficients $h_n$, and the constant $\beta_t$
is available from the authors.

{\small
\bibliographystyle{acm}
\bibliography{map3}
}

\end{document}